\documentclass[onecolumn]{aastex631}

\usepackage{natbib}
\usepackage{url}
\usepackage{amsmath}
\usepackage{CJKutf8}
\usepackage{hyperref}
\usepackage{subfigure}
\usepackage{tabularx}
\defcitealias{2025MNRAS.539.1220D}{Paper I}

\shortauthors{Dai et al.}
\begin{document}
\begin{CJK}{UTF8}{gbsn}

\title{The Samples and Binary Fractions of Red Supergiant in M31 and M33 by the HST Observations}
\correspondingauthor{Shu Wang, Biwei Jiang}
\email{shuwang@nao.cas.cn, bjiang@bnu.edu.cn}

\author[0009-0005-3951-3625]{Min Dai (代敏)}
\email{mindai@mail.bnu.edu.cn}
\affiliation{School of Physics and Astronomy, Beijing Normal University, Beijing 100875, People's Republic of China}
\affiliation{Institute for Frontiers in Astronomy and Astrophysics, Beijing Normal University, Beijing 102206, People's Republic of China}

\author[0000-0003-4489-9794]{Shu Wang (王舒)}
\affiliation{CAS Key Laboratory of Optical Astronomy, National Astronomical Observatories,
	Chinese Academy of Sciences, Beijing 100101, People's Republic of China}

\author[0000-0003-3168-2617]{Biwei Jiang (姜碧沩)}
\affiliation{School of Physics and Astronomy, Beijing Normal University, Beijing 100875, People's Republic of China}
\affiliation{Institute for Frontiers in Astronomy and Astrophysics, Beijing Normal University, Beijing 102206, People's Republic of China}

\author[0009-0001-5020-4269]{Ying Li (李颖)}
\affiliation{School of Physics and Astronomy, Beijing Normal University, Beijing 100875, People's Republic of China}
\affiliation{Institute for Frontiers in Astronomy and Astrophysics, Beijing Normal University, Beijing 102206, People's Republic of China}

\begin{abstract}
The binarity of red supergiants (RSGs) influences their evolution and the fate of supernovae. We investigate the binary fraction of RSGs in the Andromeda Galaxy (M31) and Triangulum Galaxy (M33) using photometry from the Hubble Space Telescope (HST), which offers high spatial resolution to resolve more RSGs. A preliminary step involves identifying a reliable and complete RSG sample using the F110W $-$ F160W versus F160W diagram, yielding 2,612 RSGs from the Panchromatic Hubble Andromeda Treasury (PHAT) survey of M31 3,294 RSGs from the Panchromatic Hubble Andromeda Treasury: Triangulum Extended Region (PHATTER) survey of M33. These samples suggest total RSG populations in M31 and M33 of 6,563 and 7,572, respectively. These estimates significantly exceed previous ones from the ground-based observations, an increase attributed to the superior spatial resolution of the HST. The stellar parameters of these RSGs, including effective temperature ($T_{\mathrm{eff}}$), radius ($R$), and luminosity ($L$), are derived by fitting their spectral energy distribution (SED) across optical and near-infrared bands. Binary candidates are identified by detecting ultraviolet (UV) excesses in their SEDs compared to the single-star RSG model prediction. The binary fraction is determined to be 33.4\% $\pm$ 0.9\% for M31 and 30.9\% $\pm$ 0.8\% for M33. For more luminous RSGs with log $L/L_{\odot} > 4.0$, the binary fraction decreases to 31.6\% $\pm$ 1.9\% in M31 and increases to 34.7\% $\pm$ 1.8\% in M33, respectively. These results are in good agreement with predictions from the BPASS binary evolution model. 
\end{abstract}

	
\section{Introduction} \label{sec:intro}
Binarity is prevalent among massive stars and significantly influences their evolution. Studies suggest that at least $\sim$ 50\% - 70\% of OB-type stars are binaries, with orbital periods of $0 \lesssim \log P / [\text{d}] \lesssim 3$ \citep{2008ASPC..387...93G,2012Sci...337..444S,2013A&A...550A.107S,2015A&A...580A..93D,2022A&A...665A.148S,2024arXiv241016114S}. However, the binary fraction of red supergiants (RSGs), which are descendants of OB-type main-sequence stars, seems to be much lower. Nevertheless, the binarity of RSGs influences the fate of supernovae \citep{2006ApJ...637..914W,2013A&A...558A.131G,2017A&A...601A..29Z,2018PASA...35...49E}.

Previous studies suggested that the binary fraction of RSGs ranges from $\sim$ 15\% to 40\% \citep{2019AA...624A.129P,2020AA...635A..29P,2020ApJ...900..118N,2021ApJ...908...87N,2021MNRAS.502.4890D,2022MNRAS.513.5847P}, based on various RSG samples and methodologies. Due to the severe interstellar extinction, the RSG sample in the Milky Way is highly incomplete, which limits studies of RSG binary fraction. The present results focus on the nearby galaxies, namely, the Andromeda Galaxy (M31) and the Triangulum Galaxy (M33), the Large Magellanic Cloud (LMC), and the Small Magellanic Cloud (SMC). Currently, three methods are used to identify RSG binaries: (1) variability in radial velocity, e.g., \citet{2019AA...624A.129P} found that an upper limit of 30\% on the binary fraction of RSGs in the 30 Doradus region of the LMC (a sample of 17 RSGs). \citet{2020AA...635A..29P} reported a binary fraction of $30 \pm 10\%$ in the NGC 330 region of the SMC (a sample of 15 RSGs). \citet{2021MNRAS.502.4890D} found that the minimum binary fraction is 14\% in the LMC (a sample of 51 cool supergiants) and 15\% in the SMC (a sample of 72 cool supergiants). (2) detection of far-ultraviolet (FUV) excess, e.g., \citet{2022MNRAS.513.5847P} suggested a binary fraction of $18.8\%$ in the SMC (a sample of 560 RSGs) using the Ultraviolet Imaging Telescope (UVIT) FUV (0.172 $\mu$m) photometry. (3) detection of blue-star features, e.g., \citet{2020ApJ...900..118N} combined spectra and photometry and applied machine learning to classify stars as single or binary. They found the binary fraction of RSGs to be $19.5_{-6.7}^{+7.5}\%$ in the LMC (a sample of 1,820 RSGs). \citet{2021ApJ...908...87N} found that the binary fraction of RSGs in M33 (a sample of 1,702 RSGs) is $41.2_{-7.3}^{+12}\%$ at inner regions and $15.9_{-1.9}^{+8.6}\%$ at outer regions, and for M31 (a sample of 1,909 RSGs), this fraction was $33.5_{-5.0}^{+8.6}\%$. 

We recently (\citeauthor{2025MNRAS.539.1220D} \citeyear{2025MNRAS.539.1220D}, \citetalias{2025MNRAS.539.1220D} hereafter) studied the binary fraction of RSGs in the LMC and SMC with a more complete and purer sample of RSGs, identified by using the Gaia accurate measurements of proper motion, radial velocity and parallax \citep{2021ApJ...907...18R}. Moreover, a new method was applied to identify the RSG binaries. Specifically, an object is identified as a binary system if an ultraviolet (UV) excess is detected from its spectral energy distribution (SED) fitting, as RSGs emit very weakly in the UV band. This photometry-based method is advantageous for detecting binaries with large separations and faint companions. Finally, the lower limits on the binary fraction were estimated to be 30.2\% and 32.2\% in the LMC and SMC, respectively.

In this work, we will apply this SED fitting method to study the binary fraction in M31 and M33. Given the galactic environments (e.g., metallicity) and morphologies (e.g., spiral galaxies) of M31 and M33 compared to the Magellanic Clouds (MCs), the binary fraction may be different. The difficulty is that these two galaxies are more than an order of magnitude farther than the MCs. The present largest RSG sample in M31 and M33 was obtained by \citet{2021ApJ...907...18R}, and their data are from the imaging observations by the United Kingdom Infra-Red Telescope (UKIRT), with a typical spatial resolution of 1$\arcsec$, which can bring sufficient confusion at the distance of $\sim$ 770 $-$ 800 kpc. This would lead to misclassification or missed objects. To avoid such problems, the photometric data are taken from the Hubble Space Telescope (HST) which provides superior spatial resolution, with a point spread function (PSF) full width at half maximum (FWHM) ranging from approximately 0.07$\arcsec$ to 0.13$\arcsec$, depending on the band, and a limiting magnitude of $\sim$28 mag in the F475W band. The improved spatial resolution will yield photometric accuracy comparable to that achieved in the MCs. Correspondingly, an additional step is included to identify RSGs using the HST data.

This paper is organized as follows. In Section \ref{sec:data_samples}, we present the identification of the RSG samples in M31 and M33 based on the HST photometry data. The details of fitting the SED of RSGs, identifying the RSG binaries, and calculating the binary fractions of RSGs are presented in Section \ref{sec:Method}. In Section \ref{sec:Discussions}, we compare our sample and the binary fractions with previous works, estimate the total number of RSGs in M31 and M33, assess the impact of extinction on the binary fraction, compare our results with the Binary Population and Spectral Synthesis (BPASS) model, explore the distributions of our samples in color-color diagrams, and investigate the relationship between binary fractions and metallicity. Finally, the summary is presented in Section \ref{subsec:summary}.

\section{Data and Sample} \label{sec:data_samples}
\subsection{Data}\label{datasets}
Panchromatic Hubble Andromeda Treasury (PHAT) and Panchromatic Hubble Andromeda Treasury: Triangulum Extended Region (PHATTER) are surveys by HST of M31 and M33, respectively. They both utilize the HST Wide Field Camera 3 and Advanced Camera to obtain photometry in six bands: near-ultraviolet (F275W and F336W), optical (F475W and F814W), and near-infrared (NIR) (F110W and F160W). PHAT observed roughly one-third of the high surface brightness M31 disk \citep{2023ApJS..268...48W}, and PHATTER observed the inner half of M33 \citep{2021ApJS..253...53W}.

The PHAT survey was conducted between 2010 July 12, and 2013 October 12. The most recent version of the processed catalog is V3. The PHAT observations are organized into 23 `bricks', each associated with a different proposal ID. The V3 catalog has undergone re-photometry based on the images and provides photometry for 138 million stars, achieving a limiting magnitude of $\sim$ 25 mag in F275W and F336W, $\sim$ 28 mag in F475W, $\sim$ 27 mag in F814W, $\sim$ 26.5 mag in F110W, and $\sim$ 25.5 mag in F160W \citep{2014ApJS..215....9W,2023ApJS..268...48W}. The PHATTER survey was conducted between 2017 February 21 and 2018 February 25. This catalog provides photometry for 22 million stars  in the central $\sim$ 0.1 deg$^{2}$ of M33. It reaches a limiting magnitude of $\sim$ 25 mag in F275W and F336W, $\sim$ 28 mag in F475W and F814W, $\sim$ 26.5 mag in F110W and $\sim$ 25.5 mag in F160W \citep{2021ApJS..253...53W}. These data are thus suitable for this study. In this work, we retrieved the latest V3 PHAT and PHATTER catalogs from MAST\footnote{\url{https://archive.stsci.edu/hlsp/phat}}\footnote{\url{https://archive.stsci.edu/hlsp/phatter}}. 

The initial step in the process is to screen the PHAT and PHATTER catalogs. Since RSGs are intrinsically very bright, it is possible to apply very strict photometric quality constraints without the risk of excluding genuine RSGs. This approach also minimizes the potential contamination from fainter stars with poor photometric quality. 
The specific screening criteria are as follows: F110W $\leq$ 20 mag (Tip-RGB $=$ 18.82 mag in the $J$ band, as detailed in \citeauthor{2021ApJ...907...18R} \citeyear{2021ApJ...907...18R}), the error in F814W/F110W/F160W $\leq$ 0.01 mag, the error in F475W $\leq$ 0.02 mag, and GST\_FLAG = True for PHAT, which indicates high-quality photometry. Similarly, the criteria for PHATTER are F110W $\leq$ 20.5 mag (Tip-RGB $=$ 19.2 mag in the $J$ band), the error in F814W/F110W/F160W $\leq$ 0.02 mag, the error in F475W $\leq$ 0.03 mag, and GST\_FLAG = True. After applying these criteria, the resulting catalogs contain 194,400 stars in M31 (PHAT) and 110,563 stars in M33 (PHATTER), respectively.
      
\subsection{The RSG Samples} \label{sec:Samples}
\subsubsection{Removing the Foreground Stars} \label{subsec:rm fgd}

We begin by removing foreground stars along the lines of sight toward M31 and M33. There are a few methods for this purpose. For example, Gaia astrometric data have been demonstrated to be highly effective for foreground star removal in the MCs \citep{2019A&A...629A..91Y,2020A&A...639A.116Y,2021A&A...646A.141Y}. However, due to the much larger distances to M31 and M33, only a limited number of stars have reliable astrometric measurements, and thus Gaia astrometry is used only as a complementary method. 
\citet{2016ApJ...826..224M} employed radial velocities and spectral data to remove foreground stars and identify RSGs, which is applicable to bright stars with measurable radial velocities. A more general and widely applicable method is the use of photometric color-color diagrams (CCDs), including the optical $B - V$ versus $V - R$, NIR $J - H$ versus $H - K$ and hybrid optical-infrared $z - H$ versus $r - z$ diagrams \citep{2009ApJ...703..420M,2021ApJ...907...18R,2021A&A...647A.167Y}. 
Among these, NIR CCDs offer advantages over optical ones, as RSGs (with $T_{\mathrm{eff}}$ in the range of 3,500 - 4,500 K) emit predominantly in the NIR, making them more easily detectable in the NIR than in optical bands. Furthermore, NIR bands are significantly less affected by extinction \citep{2019ApJ...877..116W, 2024ApJ...964L...3W}. 
The physical basis for using NIR CCDs to identify and remove dwarf stars is that dwarfs have higher surface gravities than giants and supergiants. This leads to more frequent atomic collisions and enhanced molecular formation, particularly in the $H$ band (1.6 $\mu$m), leading to flux suppression in dwarfs \citep{2021ApJ...907...18R}. This method was introduced by \citet{2021ApJ...907...18R,2021A&A...647A.167Y} and recently discussed in detail by \citet{2025ApJ...979..208L}.

\begin{figure*}[!htb]
	\centering
	\includegraphics[width=170mm]{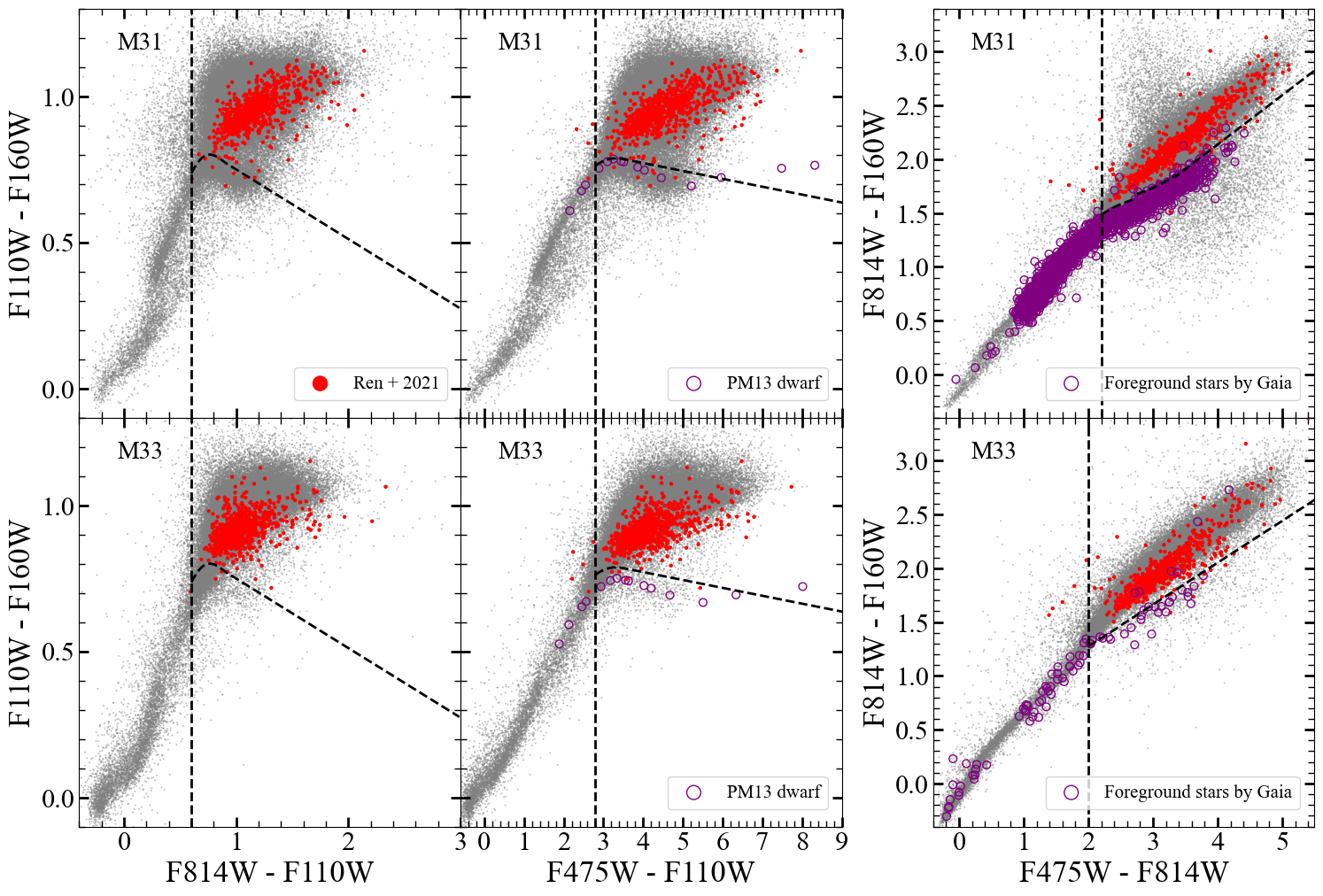}
	\caption{Three color-color diagrams are derived from HST photometry. The up panels display M31, while the down panels show M33. The red dots indicate the RSGs sample identified by \citet{2021ApJ...907...18R}. The black dashed lines represent our boundary between giant and dwarf stars.}
	\label{fig: ccds}
\end{figure*}

The PHAT and PHATTER observations provide the photometry in six bands from UV to infrared. In comparison with ground-based observations the F110W and F160W are analogous to $J$ and $H$ bands, respectively. Meanwhile, as these observations lack $K$ band coverage, the previously used NIR CCD method needed to be modified. We employ a series of CCDs to remove foreground stars, including F110W $-$ F160W versus F814W $-$ F110W, F110W $-$ F160W versus F475W $-$ F110W and F814W $-$ F160W versus F475W $-$ F814W diagrams. Additionally, Gaia/DR3 astrometry is used to remove the foreground giants. The F110W $-$ F160W versus F814W $-$ F110W diagrams for stars passing the photometric quality cuts are shown in the left panels of Figure \ref{fig: ccds} for M31 and M33. A clear bifurcation is visible in M31, where the lower-left sequence and bending branch are dwarfs, while the upper-right sequence are giants. The two branches are separated by a black dashed line, consistent with the results of \citet{2021ApJ...907...18R}. In M33, the bending branch is less pronounced due to fewer dwarf stars in the much smaller PHATTER area. The RSGs from \citet{2021ApJ...907...18R} are overplotted as red dots, which mostly fall within the giant branch region and confirm the correct separation of dwarfs and giants. A total of 14,544 and 9,816 foreground stars are independently removed by this CCD for M31 and M33, respectively. The F110W $-$ F160W versus F475W $-$ F110W diagrams are shown in the middle panels of Figure \ref{fig: ccds} for M31 and M33. The bending branch is also clearly visible in M31. The purple hollow circles indicate the theoretical dwarf locus from \citet{2013ApJS..208....9P}, providing additional confirmation that the bending branches correspond to dwarf stars. Using this CCD method, 14,092 and 11,734 foreground stars are independently removed in M31 and M33, respectively. The F814W $-$ F160W versus F475W $-$ F814W diagrams are shown in the right panels of Figure \ref{fig: ccds} for M31 and M33. The purple hollow circles in this diagram mark foreground stars identified by using Gaia DR3 astrometry. A total of 19,658 and 9509 foreground stars are independently removed by this CCD method for M31 and M33, respectively. A star is considered a foreground object and removed if it satisfies any one criterion from any of the three CCDs.

\begin{figure*}[!htb]
	\centering
	\includegraphics[width=170mm]{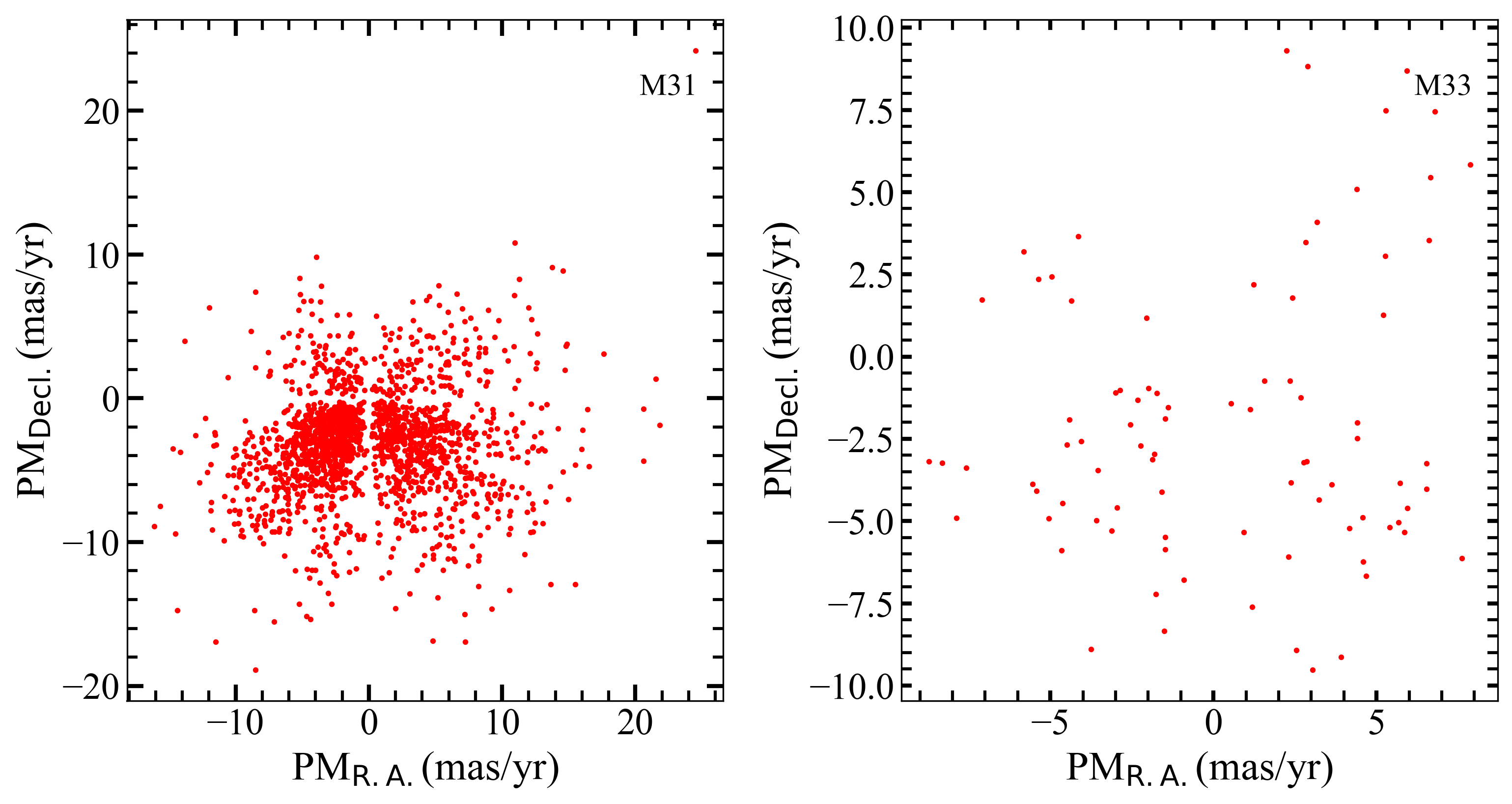}
	\caption{The distribution of proper motion for stars with reliable proper motion measurements. All stars with reliable proper motion measurements satisfy the selection criteria for foreground stars.}
	\label{fig: proper_motion}
\end{figure*}

To further improve foreground removal, we incorporate Gaia DR3 astrometry. While CCDs are effective at excluding foreground dwarfs, Gaia astrometry (parallax and proper motion measurements) is better suited for identifying foreground giants, which are expected to exhibit detectable motion. We adopt the same criteria as in \citet{2021ApJ...907...18R}, classifying stars as foreground objects if they satisfy either parallax or proper motion thresholds. We cross-match the Gaia DR3 with the PHAT and PHATTER catalogs using a matching radius of 0.18$''$, yielding 14,958 (7.7\%) matched stars in M31 and 10,146 (9.2\%) in M33. After applying aforementioned astrometric criteria, 1,905 and 86 stars are removed as foreground stars in M31 and M33, respectively. Figure \ref{fig: proper_motion} shows the distribution of proper motion for stars with reliable proper motion measurements (relative error in proper motion $<$ 20\%). Notably, due to the large distances to M31 and M33, all sources with reliable proper motion measurements satisfy the selection criteria for foreground stars (proper motion $>$ 0.2 mas$/$yr + 2$\sigma$). Among them, 8 star in M31 and 6 stars in M33 are identified as foreground stars solely by Gaia astrometry. This result confirms the finding that contamination from foreground red giants is small along the sight lines toward M31 and M33 \citep{2016ApJ...826..224M}. A detailed summary of the number of foreground stars removed by each method is provided in Table~\ref{Number of Foreground Stars}. 

\subsubsection{Identifying RSGs} \label{subsec:Identifying RSGs}
\begin{figure*}[!htb]
	\centering
	\includegraphics[width=170mm]{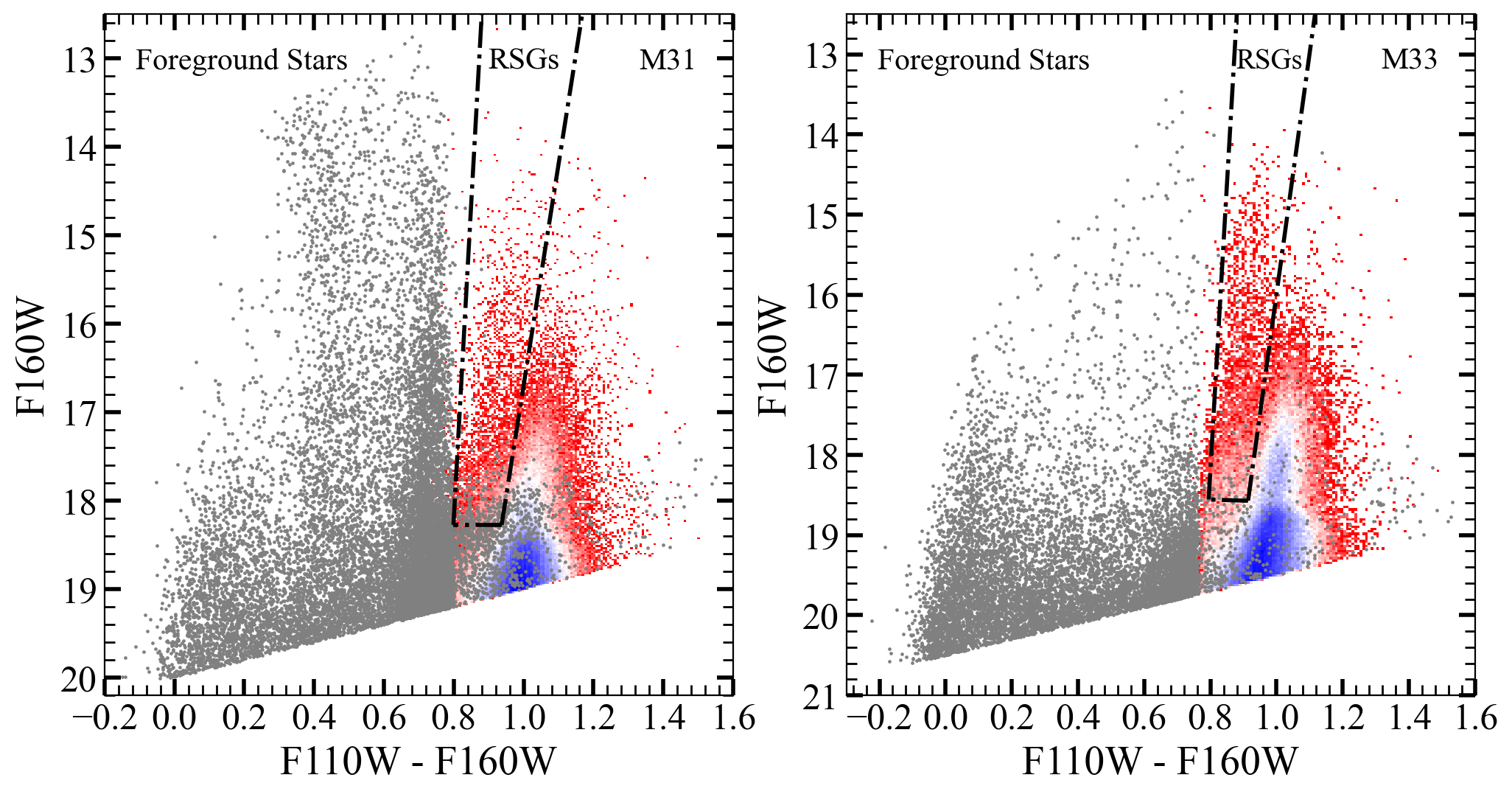}
	\caption{The observed color-magnitude diagram of the initial sample in the PHAT (left) and PHATTER (right) fields. The gray dots represent foreground stars. The member stars are colored. The black dot-dashed lines outline the RSG branches.}
	\label{fig: dwarf_giant_distribution}
\end{figure*}

After removing the foreground stars, the RSG branch becomes very prominent in the F160W versus F110W $-$ F160W diagram, as shown in Figure \ref{fig: dwarf_giant_distribution}. In M31, foreground stars (gray dots) primarily form two distinct branches at F110W $-$ F160W $\sim$ 0.42 and 0.66 mag, respectively. The RSG branch is outlined by black dot-dashed lines. A noticeable gap appears to the left of the RSG region, where only a few foreground stars fall within the RSG locus, particularly at the bright end. In contrast, contamination becomes more significant at the faint end. This trend is consistent with the predictions from the Trilegal Galactic model (e.g., see Figure 25 in \citeauthor{2021ApJS..253...53W} \citeyear{2021ApJS..253...53W}). 
The left and right boundaries of the RSG branch are visually determined based on the stellar distribution in the diagram. The mathematical expressions defining these boundaries are as follows:

\begin{equation}
	\mathrm{for\ M31 and M33},\  \mathrm{left\ boundary}: \mathrm{F160W - 18.271 = -71.250(F110W-F160W-0.799),} 
\end{equation}
\begin{equation}
	\mathrm{for\ M31},\ \mathrm{right\ boundary}: \mathrm{F160W - 18.271 = -25.000(F110W-F160W-0.937),}
\end{equation}
\begin{equation}
	\mathrm{for\ M33},\ \mathrm{right\ boundary}: \mathrm{F160W - 18.572 = -30.000(F110W-F160W-0.914).}
\end{equation}

\begin{figure*}[!htb]
	\centering
	\includegraphics[width=170mm]{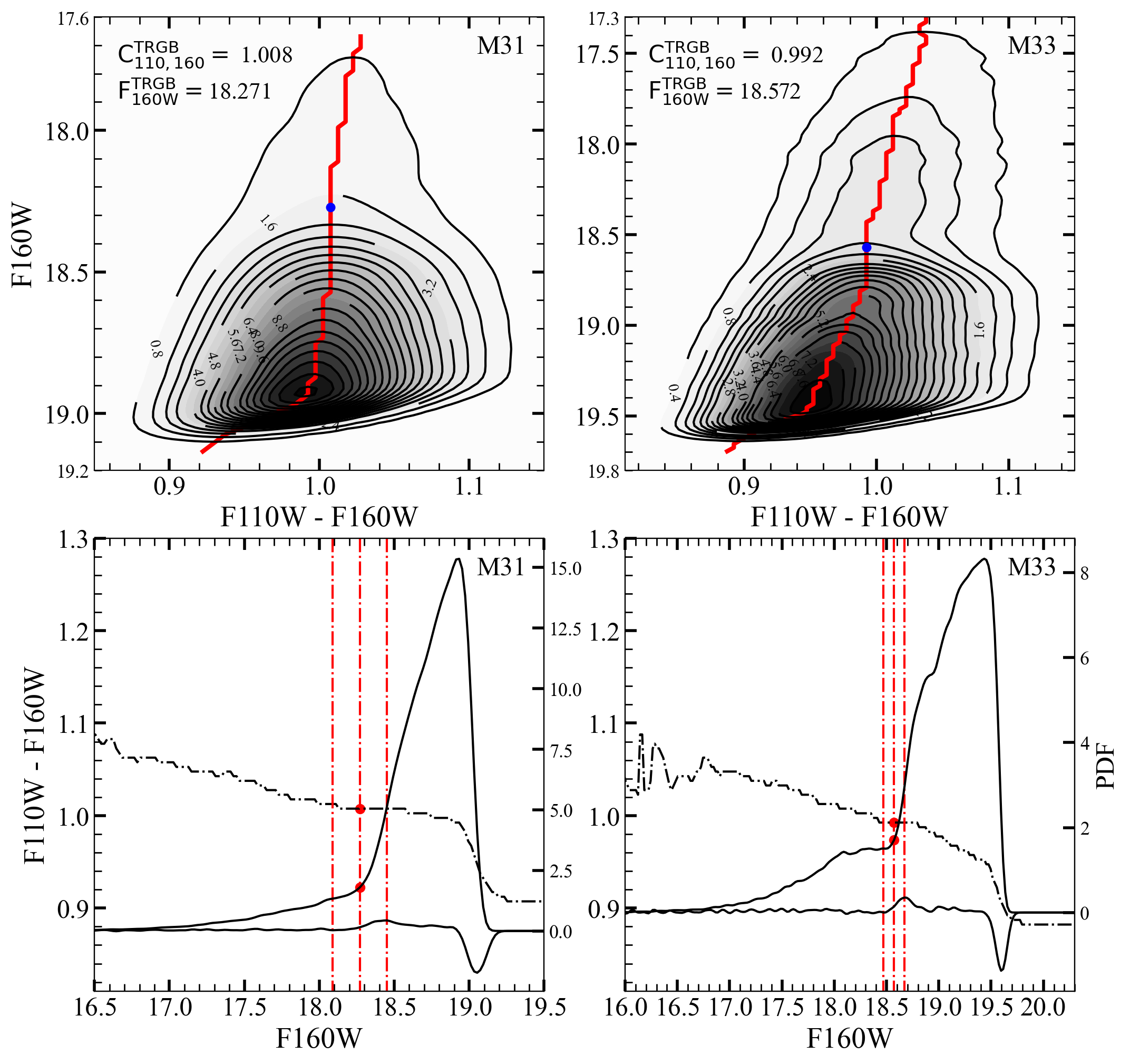}
	\caption{The probability-density function (PDF) of member stars are shown in top panels for M31 and M33, respectively, where the red line shows the F160W versus F110W $-$ F160W relation of the ridge line. The position of TRGB are shown in bottom panels, where the left and right red dashed-dotted lines shows the bright and faint ends of TRGB, respectively. The middle one shows the adopted position of TRGB. The black solid lines show the relationship between $PDF$$_{\mathrm{ridge}}$ and F160W, and the black dashed-dotted line shows the relationship between $PDF$$_{\mathrm{ridge}}$ and F110W $-$ F160W.}
	\label{fig: TRGB}
\end{figure*}

Due to the blending of faint RSGs with red giant branch (RGB) stars at the lower luminosity end, we adopt the F160W magnitude of the Tip-RGB (TRGB) as the lower luminosity boundary for RSGs, consistent with the approach of \citet{2024AJ....167..123L}. To determine the TRGB, we compute the Poisson-noise-weighted difference in star counts between adjacent bins, following the method of \citet{2018AJ....156..278G,2024AJ....167..123L}. 
We first apply Gaussian kernel density estimation using the $scikit$-$learn$ package with a bandwidth of 0.03 to derive a smoothed stellar density distribution. Poisson equation filtering is then applied to determine the precise position of the TRGB. The resulting TRGB apparent magnitudes are F160W = 18.271 mag for M31 and 18.572 mag for M33, as shown in Figure \ref{fig: TRGB}. The use of the TRGB magnitude as a reference point for delineating the lower luminosity boundary of RSGs is supported by two considerations. First, in the F160W versus F110W $-$ F160W diagram, the surface density of stars transitions from a steep decline to a more gradual slope at the TRGB location (see Figure \ref{fig: TRGB}). Second, previous studies have shown that the TRGB corresponds closely to the lower mass limit of RSGs (e.g., \citeauthor{2021ApJ...907...18R} \citeyear{2021ApJ...907...18R}). 

\begin{figure*}[!htb]
	\centering
	\includegraphics[width=170mm]{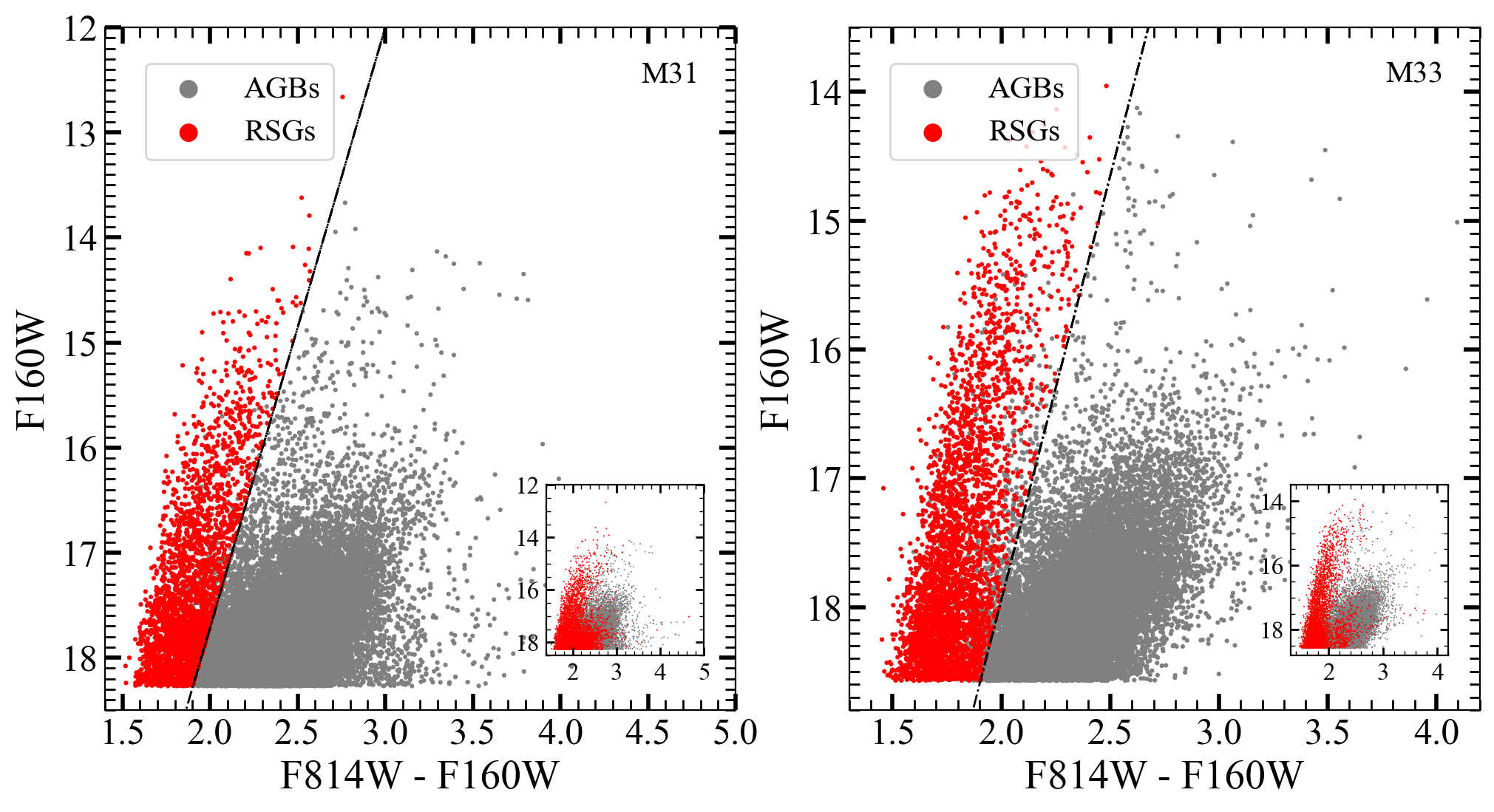}
	
	\caption{The color-magnitude diagrams for RSGs and AGBs in M31 (left) and M33 (right). The red dots represent the adopted RSG objects. The red and gray dots of inset shows the color-magnitude diagrams for objects which fall into RSG and AGB area of Figure \ref{fig: dwarf_giant_distribution}.}
	\label{fig: AGB_contaminant}
\end{figure*}

The low-luminosity portion of the RSG sample may be contaminated by asymptotic giant branch (AGB) stars \citep[e.g., see Figure 1 in][]{2024ApJ...969...81Z}. The AGB stars typically exhibit stronger stellar winds and thus have more circumstellar dust. Since the F814W band is more sensitive to dust extinction than the F110W band, AGB stars tend to appear redder in the F160W versus F814 $-$ F160W diagram. To reduce contamination from AGB stars, we further refine our RSG sample by using this diagram (Figure \ref{fig: AGB_contaminant}), removing 2,182 stars in M31 and 630 stars in M33 as likely AGB contaminants. After applying all selection criteria and removing AGB contaminants, 2,612 and 3,294 RSGs are identified in the PHAT and PHATTER surveys, respectively. 

\section{Binary fraction of RSGs} \label{sec:Method}
The method for identifying binary systems is the same as that detailed in \citetalias{2025MNRAS.539.1220D}. Due to differences in the adopted data and the objects, some parameters are adjusted accordingly. Further details of the method can be found in \citetalias{2025MNRAS.539.1220D}.
\subsection{Fitting SED of RSGs} \label{subsec:RSGs}
The observational SED of each RSG is fitted to derive its stellar parameters. A star is then identified as a binary if a significant UV excess is detected. The model spectra used for fitting are taken from the synthetic stellar spectra of Lejeune et al. (\citeyear{1997A&AS..125..229L}; hereafter L97). The L97 library covers the $T_\mathrm{eff}$ from 2,500 K to 50,000 K, log $g$ from -1.02 to 5.0, and [M/H] from -3.5 dex to 1.0 dex. 
To improve the resolution in $T_\mathrm{eff}$, the original grid spacing of 250 K is refined to 50 K through interpolation. The $\mathit{\lambda}$$\mathrm{_{eff}}$ of each filter is calculated by using the stellar parameters of a typical RSG, specifically $T_\mathrm{eff}$= 4000 K and log $g$ = 0, with [M/H] = 0.3 dex for M31, [M/H] = 0.1 dex for M33, by
\begin{equation}
	\lambda_\mathrm{eff} \equiv \frac{\int \lambda T(\lambda) S(\lambda) \, d\lambda}{\int T(\lambda) S(\lambda) \, d\lambda}\text{,}
	\label{eq:effective_wavelengths}
\end{equation}
where $T(\lambda)$ is the filter transmission curve and $S(\lambda)$ is the stellar model spectrum. The calculated $\mathit{\lambda}$$\mathrm{_{eff}}$ values are 0.2940 (0.2944), 0.3457 (0.3454), 0.4949 (0.4944), 0.8077 (0.8073), 1.1370 (1.1373), and 1.5312 (1.5320) $\mu$m in the F275W, F336W, F475W, F814W, F110W and F160W bands for [M/H] = 0.3 (0.1), respectively.

\begin{figure*}[!htb]
	\centering
	\includegraphics[width=170mm]{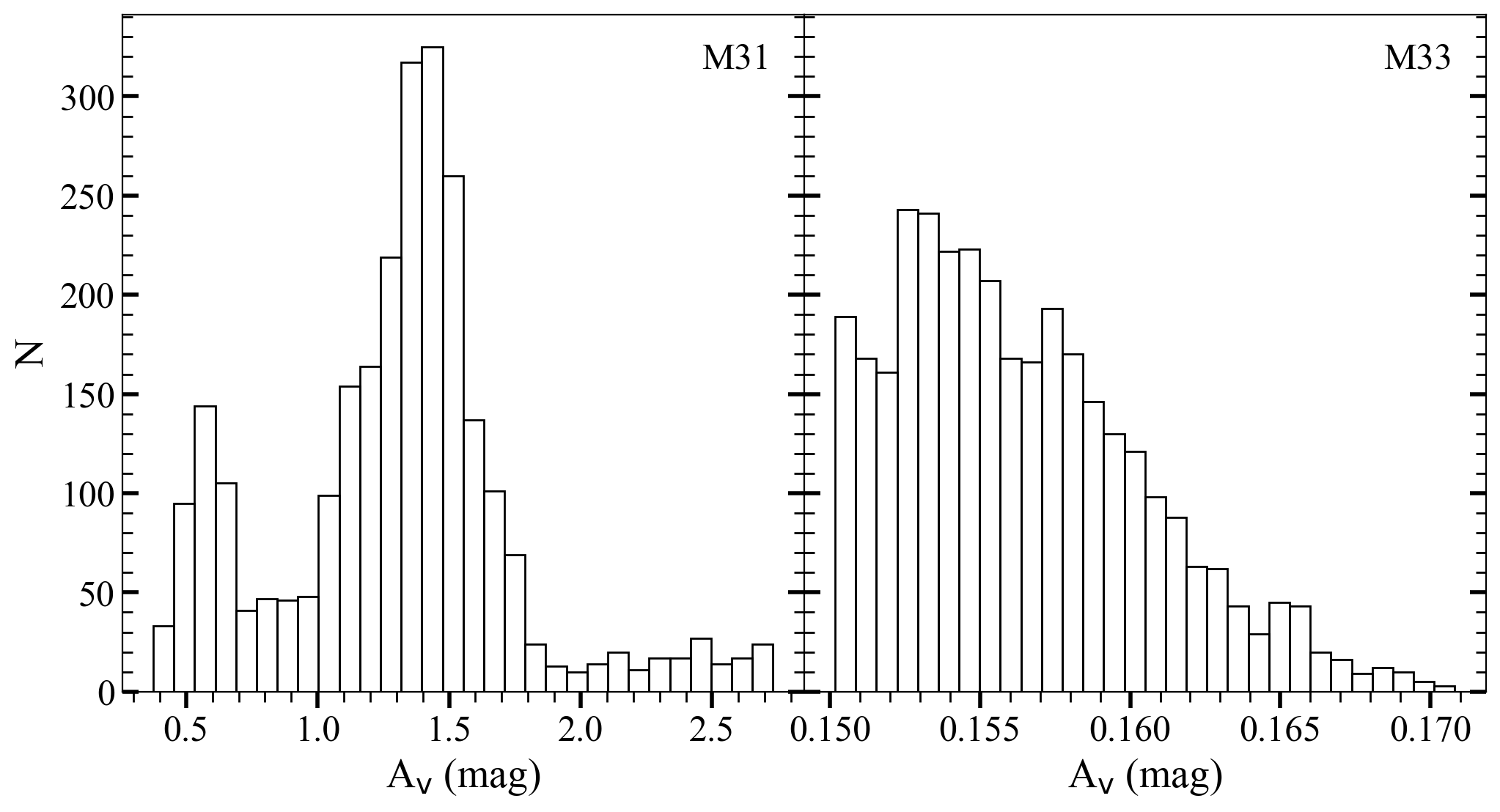}
	
	\caption{The distribution of $A_{\mathrm{V}}$ derived from SFD98 map for M31 (left) and M33 (right).}
	\label{fig: Av}
\end{figure*}

The interstellar extinction has an impact on the results of SED fitting. However, the extinction map and law for M31 and M33 are not yet fully characterized. As a reference, we use the recalibrated maps of dust infrared emission (\citeauthor{1998ApJ...500..525S} \citeyear{1998ApJ...500..525S}; \citeauthor{2011ApJ...737..103S} \citeyear{2011ApJ...737..103S}; hereafter SFD98), although it remains controversial as to whether it underestimates or overestimates extinction along certain lines of sight. Additionally, the SFD98 map is based on infrared emission that includes contributions not only from the Milky Way but also from M31 or M33 themselves. The histogram in Figure \ref{fig: Av} illustrates the $A_\mathrm{V}$ of our RSG samples, derived from the SFD98 map and converted to $A_\mathrm{V}$ using the extinction law of \citet{2019ApJ...877..116W}, which is similar to \citet{2021ApJ...912..112W}. It is important to note that adopting overly high extinction values during the SED fitting can result in unphysical estimates of $T_\mathrm{eff}$. Therefore, for M31, we initially adopt only the foreground extinction from the Milky Way, $A_\mathrm{V} = 0.17$ mag, and defer the consideration of internal extinction within M31 to later analysis (see Section~\ref{subsec:binary_fraction_of RSGs_different_extinction}). For M33, we adopt the extinction values directly from the SFD98 map. The resulting extinction values for M33 are only about a tenth of the extinction value adopted for M31. 

For the F814W, F110W and F160W bands, the fluxes are almost unaffected by contamination from companions, whose primary flux is at shorter wavelengths (e.g. $<$ 0.4 $\mu$m). However, the F475W band may be subject to such contamination. Our SED fitting strategy therefore proceeds in two steps. First, we fit the SED for all RSGs using only the F814W, F110W and F160W bands. For a subset of stars (99 in M31 and 88 in M33) where the observed F475W flux is significantly lower than the model-predicted flux, we then perform a second round of fitting using all four bands: F475W, F814W, F110W and F160W. To evaluate the goodness of fit, the reduced $\chi^2$ (same as \citetalias{2025MNRAS.539.1220D}) for each star is calculated as
\begin{equation}
\begin{split}
 \frac{\chi^2}{\mathrm{dof}}=\frac{\sum_{j=1}^{N_{\mathrm{obs}}}[F_\nu(\lambda_j)^{\mathrm{mod}}-F_\nu(\lambda_j)^{\mathrm{obs}}]^2/w(\lambda_j)^2}{N_{\mathrm{obs}}-N_{\mathrm{para}}}\text{,}
\end{split}
\label{eq:chi-square}
\end{equation}
 where $F_\nu(\lambda_j)^{\mathrm{mod}}$ and $F_\nu(\lambda_j)^{\mathrm{obs}}$ are the model and the observed flux after correcting the extinction, respectively. $F_\nu(\lambda_j)^{\mathrm{mod}}$ is calculated by
\begin{equation}
F_\nu(\lambda_j)^{\mathrm{mod}} \equiv \frac{\int T(\lambda) S(\lambda) \, d\lambda}{\int T(\lambda) \, d\lambda}\text{.}
\label{eq:mod_flux}
\end{equation}
The weight is defined as $w(\lambda_j) = [F_\nu(\lambda_j)^{\mathrm{mod}} + F_\nu(\lambda_j)^{\mathrm{obs}}]/2$ to eliminate the effect caused by the large flux differences between various bands, across very wide wavelength range. $N_\mathrm{obs}$ is the number of observations and $N_\mathrm{para}$ is the number of free model parameters. We only fit for $T_\mathrm{eff}$. While, log $g$ is set to 0\footnote{Due to model limitations, log $g$ = 0.28 for $T_\mathrm{eff}$ $<$ 3500 K.}, and [M/H] are set to 0.3 dex for M31 or 0.1 dex for M33. This is done because no reliable measurements of log $g$ and [M/H] are available for individual objects, and the dispersion in these two parameters is expected to be small for the RSG population. Using the fitted $T_\mathrm{eff}$, the stellar radius ($R$) and luminosity ($L$) are calculated by
\begin{equation}
    R = d \sqrt{\frac{F_\nu(\lambda)^{\mathrm{obs}}}{F_\nu(\lambda)^{\mathrm{mod}}}}\text{,}
\label{eq:cal_R}
\end{equation}
\begin{equation}
	L = 4\pi R^2 F\text{,}
	\label{eq:cal_L}
\end{equation}
where d is the distance to M31 or M33, and $F$ is the integrated flux of the model spectrum. Distance moduli of 24.47 mag for M31 and 24.54 mag for M33 are adopted to calculate the $d$ \citep{2005MNRAS.356..979M,2012AJ....144....4M}. Tables \ref{tab: M31_results} and \ref{tab: M33_results} list the identification numbers of RSGs identified in this work, their physical parameters derived from the SED fitting, and their classification for RSGs in M31 and M33, respectively. 

\begin{figure*}[!htb]
	\centering
	\includegraphics[width=170mm]{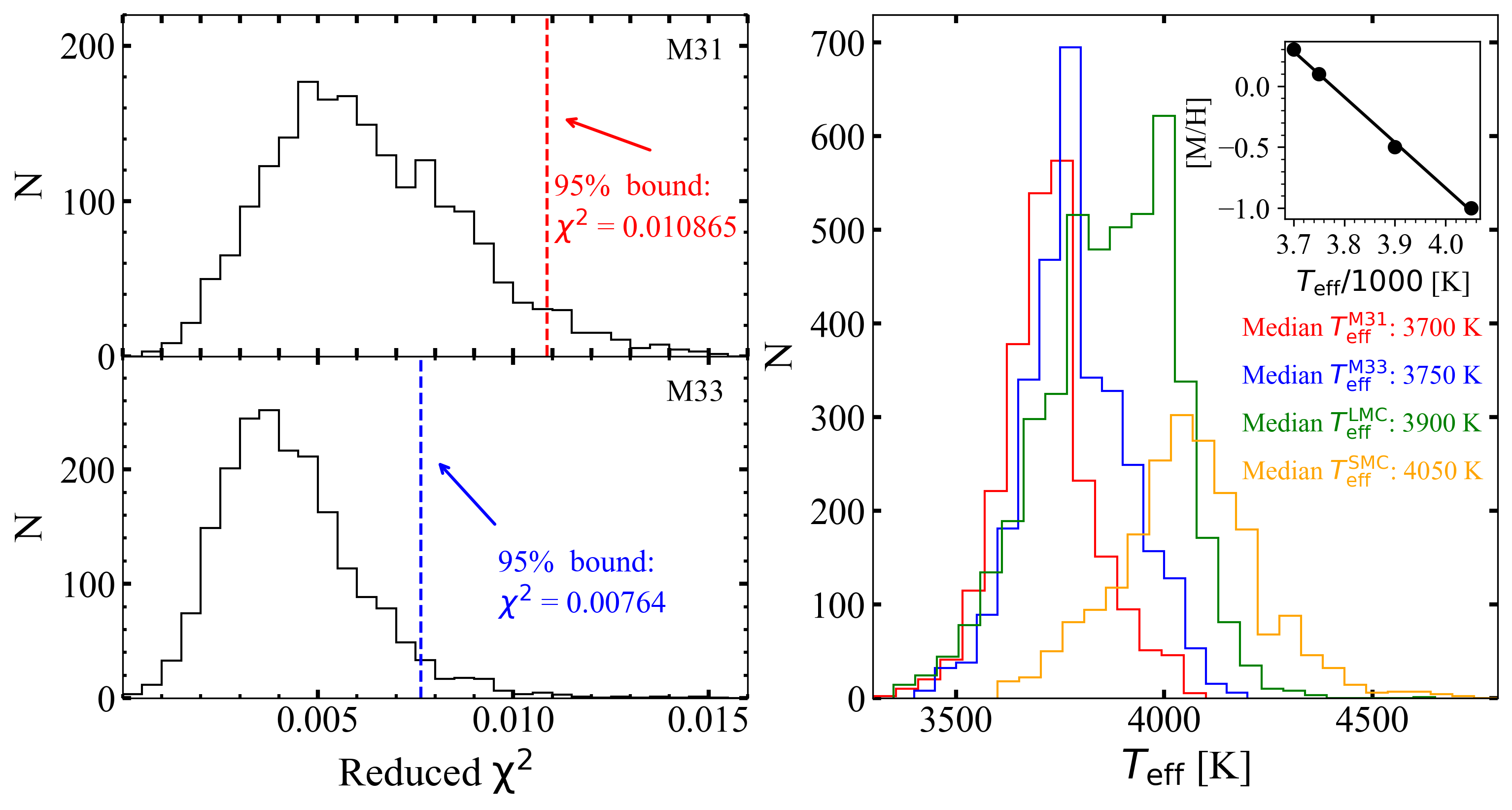}
	\caption{Left panel shows the distribution of reduced $\chi^2$ derived from SED fitting for M31 and M33, where the red and blue dashed lines represent the 95\% confidence interval of our sample. Right panel shows the distribution of $T_{\mathrm{eff}}$ derived from SED fitting for M31, M33, LMC and SMC, where the inset shows relationship between [M/H] and median $T_{\mathrm{eff}}$ of RSGs in the four galaxies mentioned above.}
	\label{fig: teff_chi_distribution}
\end{figure*}

The distributions of reduced $\chi^2$ and $T_{\mathrm{eff}}$ from the SED fitting of RSGs are presented in Figure \ref{fig: teff_chi_distribution}. In the left panel, red and blue dashed lines denote the critical values of $\chi^2$ = 0.010865 and 0.007640 at the 95\% confidence level for M31 and M33, respectively. The 131 RSGs in M31 and 165 in M33 with $\chi^2$ values exceeding the critical limits are excluded due to poor fitting.
The right panel shows the $T_{\mathrm{eff}}$ distributions for RSGs in M31 and M33. For comparison, the LMC and the SMC are also included in this panel. The median $T_{\mathrm{eff}}$ values for the LMC and SMC are derived from \citetalias{2025MNRAS.539.1220D}, where the corresponding [M/H] values are set to -0.5 dex and -1.0 dex, respectively. Overall, the median $T_{\mathrm{eff}}$ values for M31 (3,700 K) and M33 (3,750 K) are lower than those for the LMC (3,900 K) and SMC (4,050 K). The trend of increasing $T_{\mathrm{eff}}$ with decreasing [M/H] among RSGs can be attributed to the effect of metallicity on atmospheric opacity: higher metallicity increases opacity, thereby reducing the effective temperature. A strong linear correlation is found between [M/H] and $T_{\mathrm{eff}}$. The black dots in the inset of the right panel represent the [M/H] and median $T_{\mathrm{eff}}$ values for the four galaxies, while the black line represents the linear fit: $\mathrm{[M/H]} = -3.733(\pm 0.120)T_{\mathrm{eff}}/1000 + 14.098(\pm 0.463)$.

\subsection{Identification of RSG Binary} \label{subsec:Binary_fraction}
The criterion for identifying an RSG as a binary is based on the comparison between $F_\nu(\lambda_j)^{\mathrm{mod}}$ and $F_\nu(\lambda_j)^{\mathrm{obs}}$ in the UV bands. Specifically, when the observed UV flux exceeds the model prediction by more than 3$\sigma$, the star is identified as a binary system. Our binary identification criterion is expressed as
\begin{equation}
	F_\nu(\lambda_j)^{\mathrm{obs}} - F_\nu(\lambda_j)^{\mathrm{mod}}> 3 F^{{\mathrm{err}}}_{\text{\(\nu\)}}(\lambda_j)^{\mathrm{obs}}  + 3 F^{{\mathrm{err}}}_{\text{\(\nu\)}}(\lambda_j)^{\mathrm{mod}} \text{,}
	\label{eq:binary_criterion}
\end{equation}
where $F^{{\mathrm{err}}}_{\text{\(\nu\)}}(\lambda_j)^{\mathrm{obs}}$ is the observational error, and $\lambda_j$ is F275W or F336W when F275W is unavailable. Literature suggests that the log $g$ of an RSG can be up to 0.5 \citep{2017ars..book.....L}. Accounting for the influence of log $g$, where a higher log $g$ for an RSG results in elevated UV model fluxes, we define $F_{\nu}^{\mathrm{err}}(\lambda_j)^{\mathrm{mod}}$ as the error in the model flux,
\begin{equation}
	F_{\nu}^{\mathrm{err}}(\lambda_j)^{\mathrm{mod}} = F_\nu^{\mathrm{max}}(\lambda_j)^{\mathrm{mod}} - F_\nu(\lambda_j)^{\mathrm{mod}}  \text{,}
\end{equation}
where $F_\nu^{\mathrm{max}}(\lambda_j)^{\mathrm{mod}}$ is the model flux with log $g$ = 0.5\footnote{Due to model limitations, log $g$ = 0.6 for $T_\mathrm{eff}$ $<$ 3500 K.} and $T_{\rm eff}$ = $T_{\rm eff}^{\mathrm{RSG}}$. The $T_{\rm eff}^{\mathrm{RSG}}$ and $F_\nu(\lambda_j)^{\mathrm{mod}}$ are the $T_{\rm eff}$ and model flux of the RSG derived from SED fitting (see Section \ref{subsec:RSGs}). For an example star with typical photometric error (0.002, 0.002 and 0.001 mag in F814W, F110W and F160W, respectively), we perform a total of 2,000 SED fittings to the observational SEDs that are generated from a Gaussian distribution based on the photometric flux and its error. These fittings yield a single value $T_{\mathrm{eff}}$, which implies that the fitted $T_{\mathrm{eff}}$ error is smaller than the $T_{\mathrm{eff}}$ grid interval of the model (50 K). Therefore, we only consider the uncertainties introduced by variations in log $g$ and neglect the errors in the fitted $T_{\mathrm{eff}}$. 

\begin{figure*}[!htb]
	\centering
	\includegraphics[width=170mm]{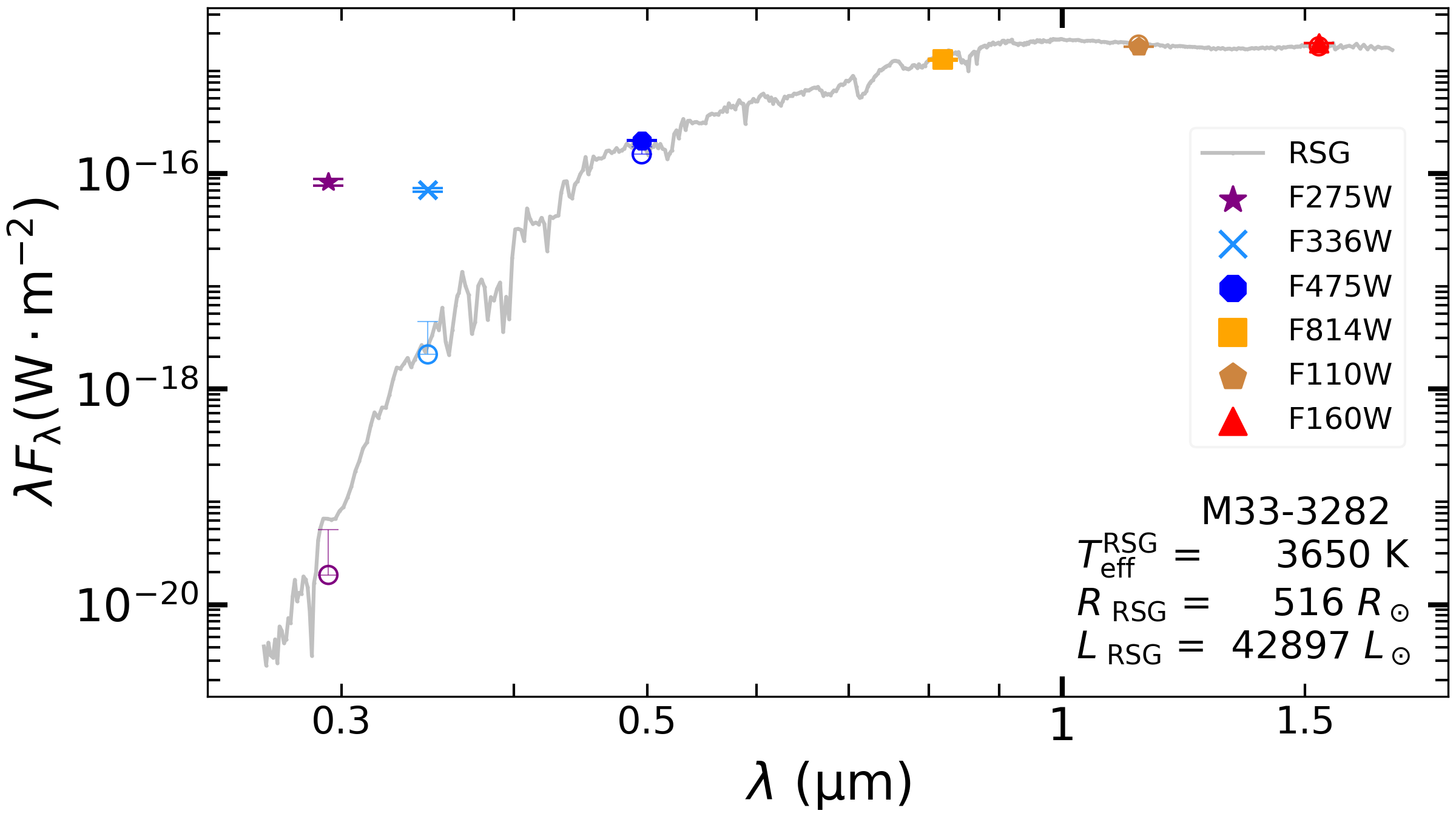}
	\caption{An example of the SED fitting to the RSG component, M33-3282. The gray solid line represents the L97 model spectrum. The solid and hollow markers with 3$\sigma$ error bars show the photometry and model data. The No., $T_{\mathrm{eff}}$, $R$ and $L$ of object are displayed on the lower right.}
	\label{RSG SED fitting}
\end{figure*}

An example of the SED fitting for an RSG binary is shown in Figure \ref{RSG SED fitting}, where the observed fluxes in the F275W and F336W bands are significantly higher than the RSG model spectrum (gray solid line), and the star is identified as a binary with a hot companion. Finally, 828 and 966 binary candidates are identified in M31 and M33, respectively. The binary fraction of RSGs is then 33.4\% $\pm$ 0.9\% (828/2481) in M31 and 30.9\% $\pm$ 0.8\% (966/3129) in M33, respectively, where the fraction uncertainties are estimated based on a binomial distribution.

\section{Discussion} \label{sec:Discussions}
\subsection{Comparison with Previous Works on the RSG Sample} \label{subsec:Comparison_samples}

\begin{figure*}[!htb]
    \centering
    \includegraphics[width=170mm]{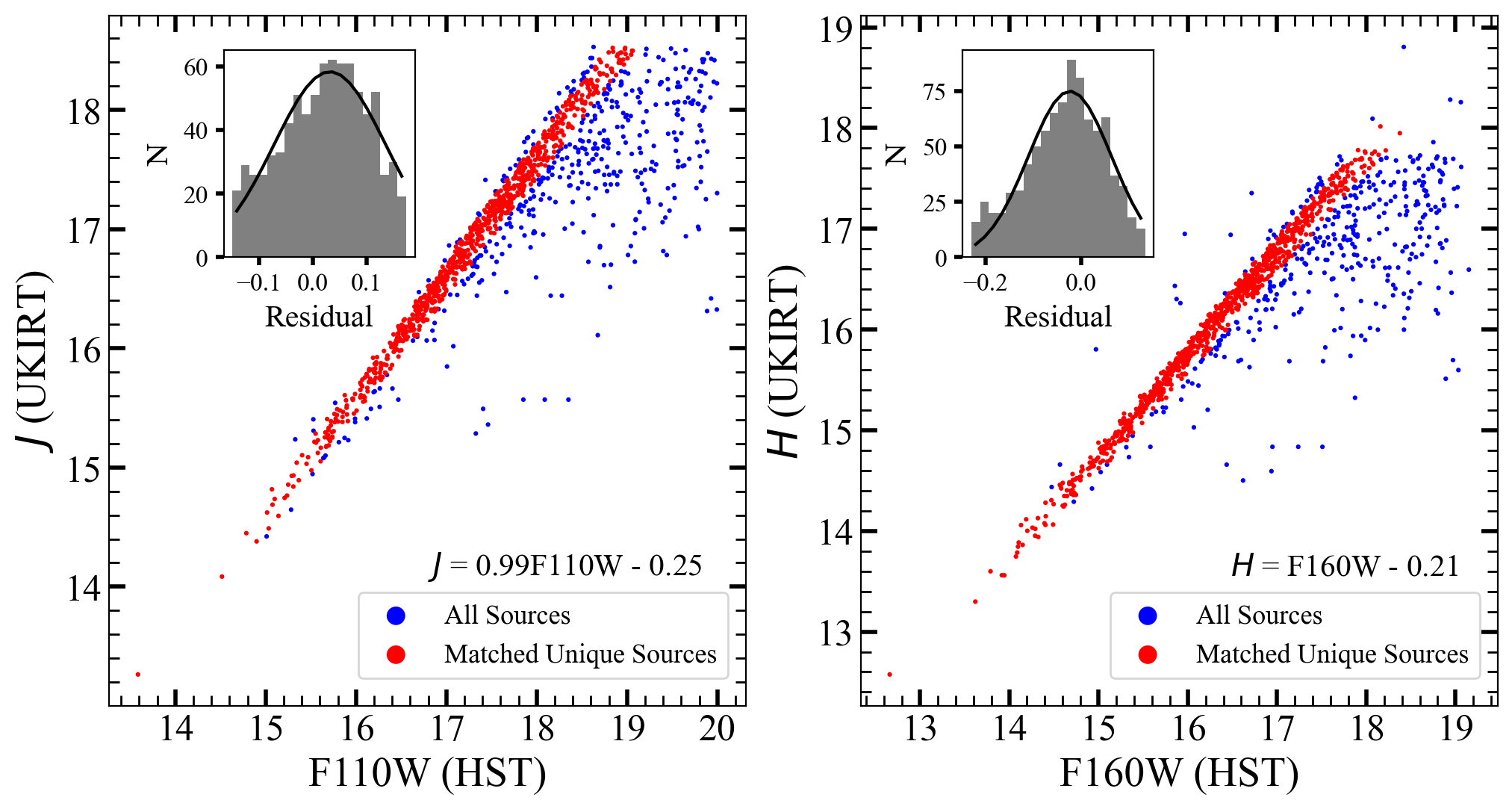} 
    \caption{Comparison of HST/F110W and F160W photometry with matched UKIRT/$J$ and $H$ photometry from \citet{2021ApJ...907...18R} in M31. The red dots represent the matched unique sources. The blue dots mark the UKIRT sources that matched more than one HST source.}
    \label{fig: magnitudes_comparison}
\end{figure*}

There are 5,498 and 3,055 RSGs identified by \citet{2021ApJ...907...18R} in M31 and M33, respectively. Among these, 2,186 (39.8\%) and 1,330 (43.5\%) RSGs are located within the PHAT and PHATTER regions, respectively. A radius of 0.5$''$ is adopted to cross-match the RSG samples of \citet{2021ApJ...907...18R} with the PHAT and PHATTER catalogs. However, due to the resolution and photometric limitations of ground-based telescopes, a fraction of sources exhibit inconsistent magnitudes (e.g. F110W versus $J$ and F160W versus $H$) as shown in Figure \ref{fig: magnitudes_comparison}. We apply a 3$\sigma$ clipping method to establish reliable one-to-one relationships between these photometric systems. In Figure \ref{fig: magnitudes_comparison}, the red dots represent the matched RSG pairs, while the blue dots denote unmatched sources. Finally, 782 and 883 matched pairs are found in the PHAT and PHATTER, respectively. Based on these pairs, we derive empirical conversion relations between the UKIRT and HST bands: 
\begin{itemize} 
\item $J$ = 0.99($\pm$0.004)F110W - 0.25($\pm$0.062), $H$ = 1.0($\pm$0.004)F160W - 0.21($\pm$0.058) for PHAT,  
\item $J$ = 1.00($\pm$0.003)F110W - 0.45($\pm$0.058), $H$ = 1.0($\pm$0.003)F160W - 0.28($\pm$0.053) for PHATTER.
\end{itemize}
The unmatched stars missing from the ground-based telescope observations do so because they remain unresolved at the ground-based resolution (see also \citeauthor{2012AJ....143...74B} \citeyear{2012AJ....143...74B}). 
 
\begin{figure*}[!htb]
	\centering
	\includegraphics[width=170mm, height=84mm]{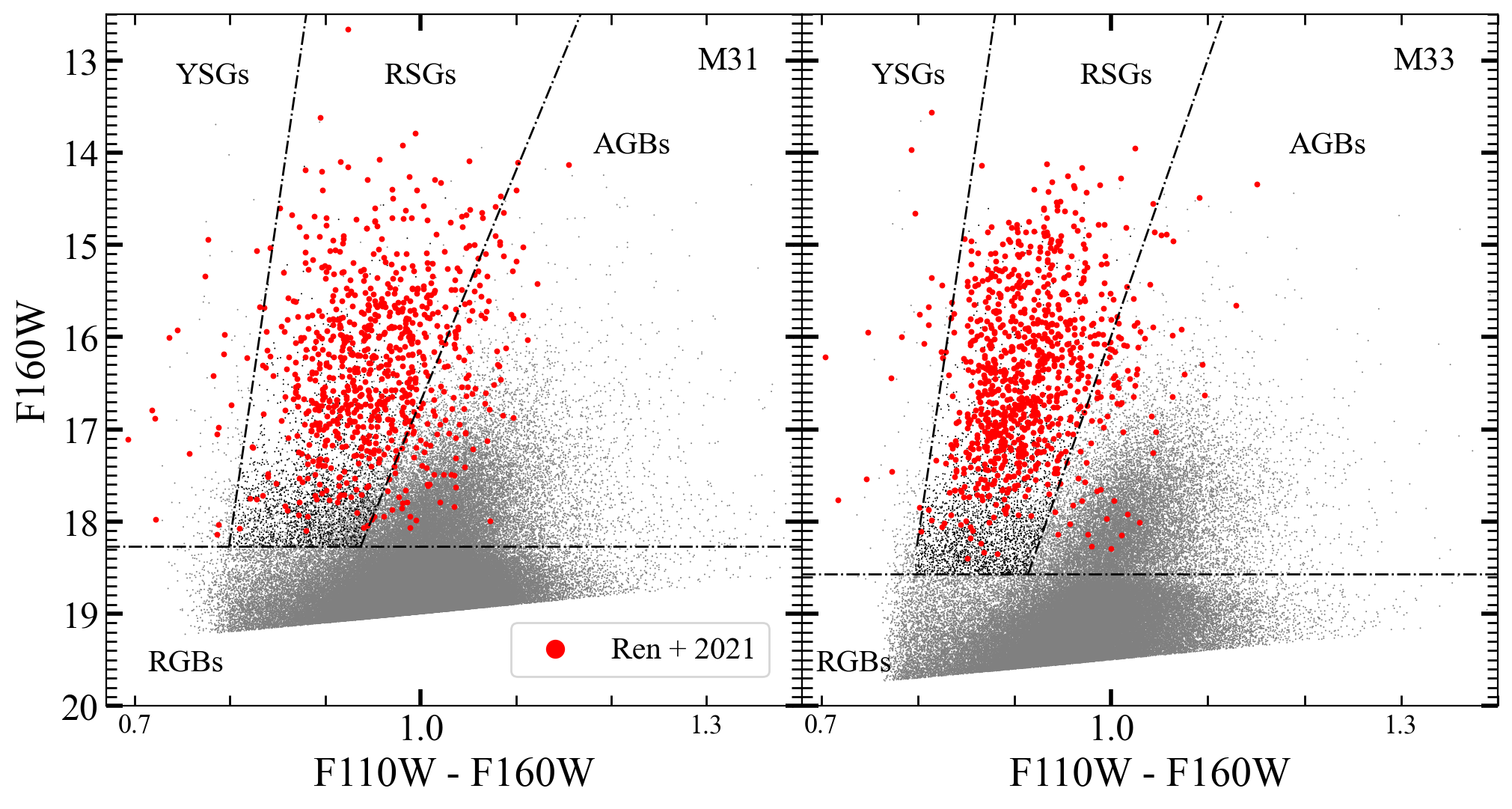}
	\caption{The detailed population classification for member stars based on color-magnitude diagrams. The cross-matched RSGs sample from \citet{2021ApJ...907...18R} are marked as red dots.}
	\label{fig: cmd}
\end{figure*}

Figure \ref{fig: cmd} compares the positions of our RSGs with those from \citet{2021ApJ...907...18R} (red dots) in the F160W versus F110W $-$ F160W diagrams. Most of the red dots fall well within our defined RSG region, demonstrating the validity of our RSG selection. Nevertheless, our data reveal many additional faint RSGs, and some RSGs identified by \citet{2021ApJ...907...18R} lie outside our defined RSGs region, towards either its bluer or redder side. In addition, we identify 147,877 and 77,309 RGB stars, 22,679 and 17,159 AGB stars, and 212 and 227 candidate yellow supergiants (YSGs) in M31 and M33, respectively.

\subsection{Distribution and Total Number of RSGs in M31 and M33.} \label{subsec:Estimating}
\begin{figure*}[!htb]
	\centering
	\subfigure{\includegraphics[width=84mm, height=84mm]{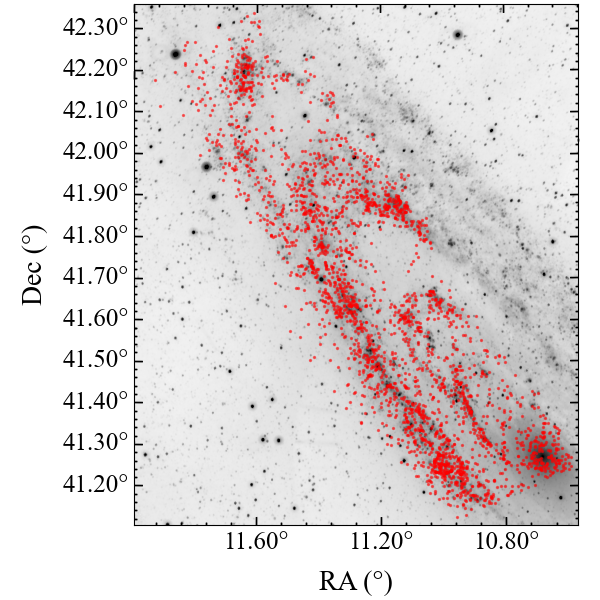}}\hspace{2mm}
	\subfigure{\includegraphics[width=84mm, height=84mm]{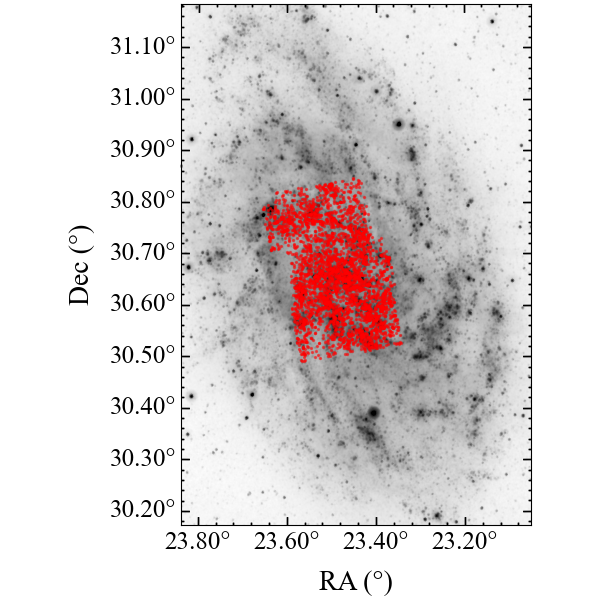}}
	\caption{Spatial distribution of RSGs in M31 (left) and M33 (right). The background image is taken from the GALEX NUV observation.}
	\label{fig: rsg_distribution}
\end{figure*}

Figure \ref{fig: rsg_distribution} illustrates the spatial distribution of our RSGs (indicated by red dots) in M31 and M33, overlaid on the GALEX NUV images. In both galaxies, the RSGs are clearly concentrated in the spiral arms. This is consistent with RSGs being massive stars typically found in star-forming regions, thus further supporting the reliability of our RSG identification.

The total number of RSGs in M31 and M33 is estimated to be approximately 6,563 and 7,572, respectively. This estimation is achieved by scaling our sample sizes from the PHAT and PHATTER regions, using the fractions of the total RSG populations (39.8$\%$ in M31 for PHAT and 43.5$\%$ in M33 for PHATTER) that these regions are understood from \citet{2021ApJ...907...18R} to contain relative to the entire galaxy. In comparison, \citet{2021ApJ...907...18R} identified 5,498 and 3,055 RSGs in M31 and M33, respectively. Thus, the present work identified 19\% and 148\% in M31 and M33 respectively more RSGs than ground-based observations. This increase is attributed to the superior spatial resolution of HST.

\subsection{Binary Fraction of RSGs with log $L/L_{\odot} > 4.0$} \label{subsec:binary_fraction_of RSGs_with_L>4.0}
The low-luminosity RSG sample may be contaminated by AGB stars (e.g., see Figure \ref{fig: AGB_contaminant}). To address this, we apply a luminosity threshold of log $L/L_{\odot} > 4.0$ to select a high-luminosity subsample of RSGs. This subsample consists of 584 RSGs in M31 and 735 RSGs in M33. The corresponding binary fractions become 31.6\% $\pm$ 1.9\% (185/585) in M31 and 34.7\% $\pm$ 1.8\% (254/733) in M33. Relative to the full sample, the binary fraction in this subsample decreases by 1.8\% in M31 and increases by 3.8\% in M33. 

\subsection{Binary Fraction of RSGs by Accounting Different Extinction} \label{subsec:binary_fraction_of RSGs_different_extinction}
As mentioned in Section \ref{subsec:RSGs}, we discussed the impact of varying internal extinction assumptions in M31 on the results of SED fitting and the subsequently derived binary fraction. Specifically, we tested three extinction scenarios: (1) the Milky Way foreground extinction combined with the full SFD98 extinction, (2) the foreground extinction plus half of the SFD98 extinction, and (3) the foreground extinction plus one-third of the SFD98 extinction. The corresponding binary fractions are 32.7\% (811/2,481), 36.9\% (916/2,481), and 36.5\% (906/2,481), respectively. 

\begin{figure*}[!htb]
	\centering
	\includegraphics[width=85mm]{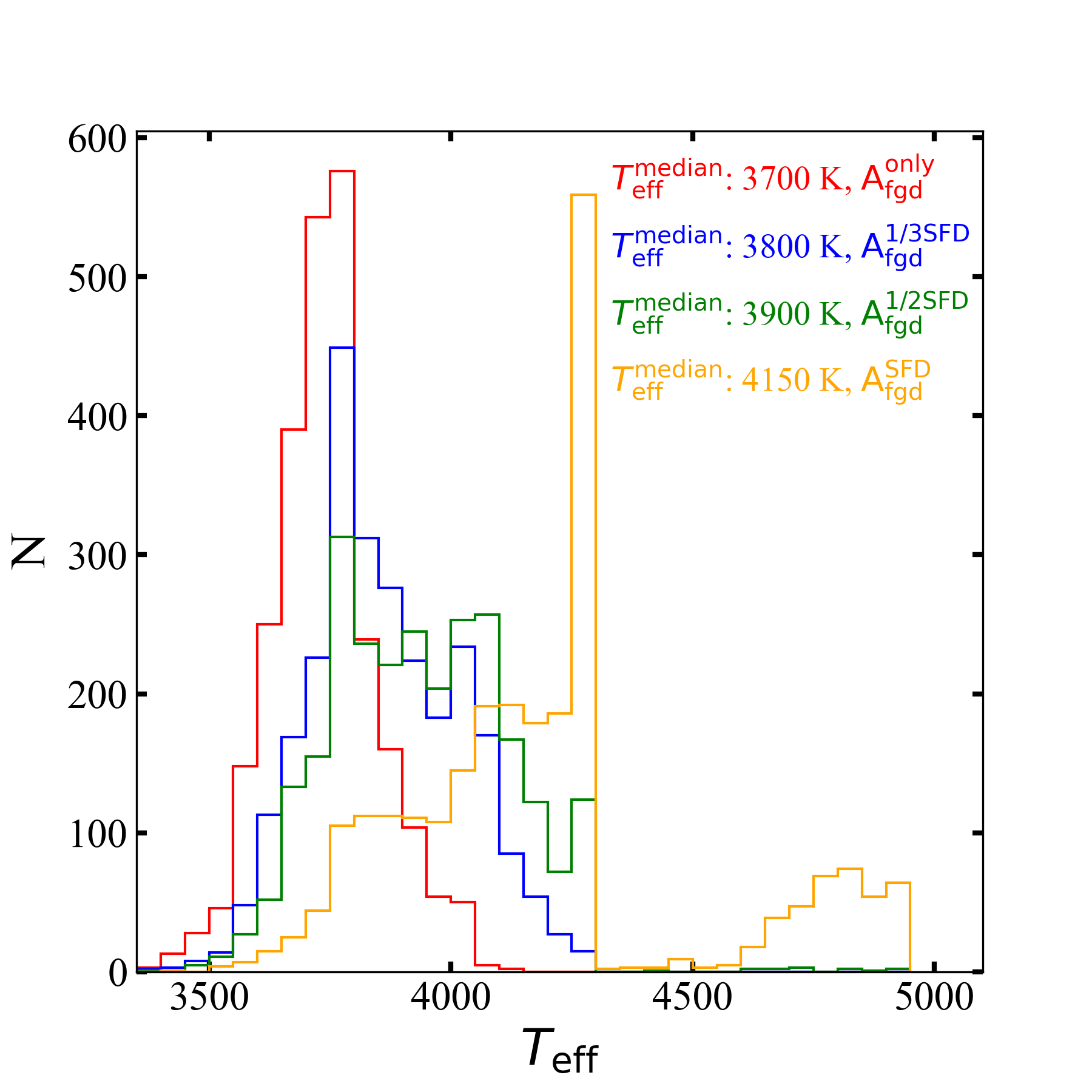}
	\caption{The distribution of $T_{\mathrm{eff}}$ derived from SED fitting based on several SFD extinction considerations in M31. `A$_{\mathrm{fgd}}$' refers to the foreground extinction of the Milky Way, adopting an $A_{\mathrm{V}}$ value of 0.17 mag for M31.}
	\label{fig: different_extinction}
\end{figure*}

Figure~\ref{fig: different_extinction} shows the $T_{\mathrm{eff}}$ distributions obtained from SED fitting under these extinction scenarios. Theoretical relationship predict an inverse correlation between [M/H] and $T_{\mathrm{eff}}$ for RSGs, where higher [M/H] leads to lower $T_{\mathrm{eff}}$. Given this expectation, the scenario using only the Milky Way foreground extinction yields the most physically consistent results, as illustrated in Figure \ref{fig: teff_chi_distribution}. Although correcting for extinction strongly enhances the UV flux, it also increases the fitted $T_{\mathrm{eff}}$, which in turn reduces its overall impact on the inferred binary fraction. As a result, the variation in binary fraction across the different extinction assumptions remains relatively small. 

\subsection{Comparison with the BPASS Model on the Fraction of Binary} \label{subsec:Comparison_binary_fraction_model}
BPASS is a widely utilized framework for stellar population evolution and spectral synthesis. It incorporates evolutionary models with diverse parameters, such as metallicity and initial mass function (IMF), to simulate the physical properties and spectral characteristics of stellar populations \citep{2018MNRAS.479...75S}. We retrieved the pre-compiled data\footnote{\url{https://zenodo.org/records/4064300}} with the fiducial IMF and a Kroupa maximum mass of 300 $M_{\odot}$, using the $hoki$\footnote{\url{https://github.com/HeloiseS/hoki}} package. To select appropriate RSGs models from BPASS for comparison with our observational data, we applied the following criteria: 
\begin{align}
    \text{For M31 and M33:} & \quad M_{\mathrm{ini}} \geq 8 \mathrm{M_{\odot}}, \ 8 \mathrm{M_{\odot}} \leq M_{\mathrm{c}} \leq 30 \mathrm{M_{\odot}}, \ \log L/L_{\odot} \geq 4.0, \\
    \text{For M31:} & \quad 3350 \mathrm{K} \leq T_{\mathrm{eff}} \leq 4050 \mathrm{K}, \ \mathrm{Z} = 0.04,\\
	\text{For M33:} & \quad 3450 \mathrm{K} \leq T_{\mathrm{eff}} \leq 4100 \mathrm{K}, \ \mathrm{Z} = 0.02,
\end{align}
where $M_{\mathrm{ini}}$ is the initial mass, $M_{\mathrm{c}}$ is the current mass, Z is the metallicity mass fraction. The conversion relation between Z and [M/H] is Z = Z$_{\odot}\cdot$10$^{\mathrm{[M/H]}}$, where Z$_{\odot}$ = 0.02 represents the solar metallicity \citep{2018MNRAS.479...75S}.
The $T_{\mathrm{eff}}$ range is derived from our observational results presented in Figure \ref{fig: teff_chi_distribution}. 

Using the above criteria, the BPASS model predicts RSGs binary fractions of 37\% in M31 and 46\% in M33. However, the BPASS binary population includes a number of systems with low-luminosity main-sequence companions that have not yet reached the zero-age main sequence (ZAMS) by the time RSGs form. These companions are not expected to be observable in observational samples and should therefore be excluded from comparison with observations. Therefore, we refine our criteria by selecting only binary systems with luminous OB-type main-sequence companions, defined as having $L/L_{\odot} \geq 60.7$, i.e., the luminosity of a B9V star. Under these revised criteria, the predicted binary fractions from the BPASS model decrease to 30.9\% for M31 and to 38.6\% for M33. These results are in good agreement with the binary fractions derived from our observational subsample of RSGs with log $L/L_{\odot} > 4.0$, which are 31.6\% for M31 and 34.7\% for M33. 

\subsection{Comparison with Previous Work on the Fraction of Binary} \label{subsec:Comparison_binary_fraction}
\citet{2021ApJ...908...87N} trained a machine learning model using a combination of spectroscopic and photometric data, subsequently applying it to multi-band photometric data to calculate the binary fraction of RSGs. They reported a final binary fraction in M31 of 33.5\%, which is in excellent agreement with our results of 33.4\% for the full sample and 31.6\% for the subsample with log $L/L_{\odot} > 4.0$. Meanwhile, for M33, they found a clear metallicity-dependent trend in the binary fraction: 15.9\% in the outer region, 26.9\% in the middle region, and 41.2\% in the inner region. Considering that the PHATTER survey covers primarily the middle and inner regions of M33, our results (30.9\% for the full sample and 34.7\% for the subsample with log $L/L_{\odot} > 4.0$) are consistent with those of \citet{2021ApJ...908...87N}.

\begin{figure*}[!htb]
	\centering
	\subfigure{\includegraphics[height=160mm]{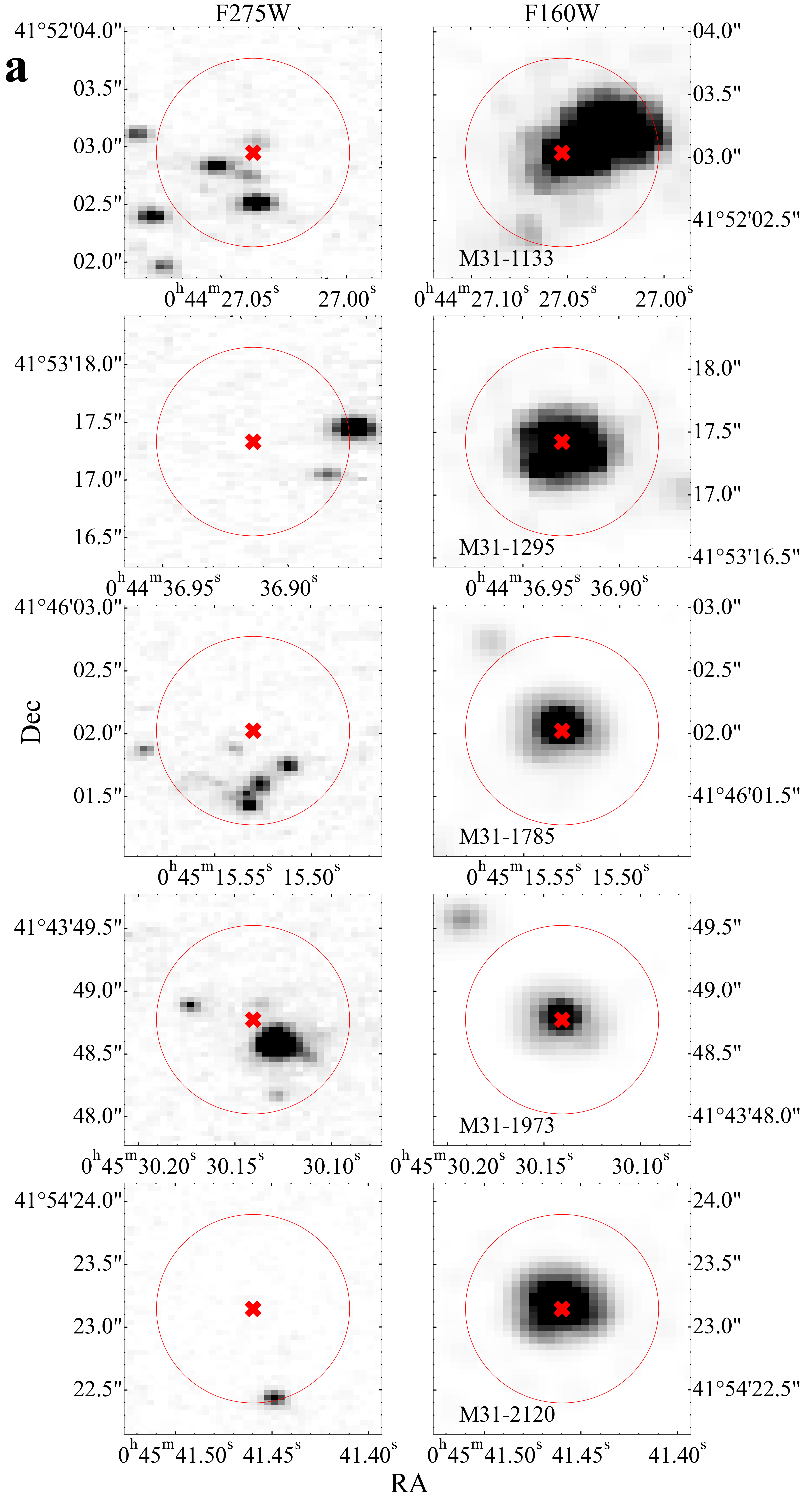}}
	\subfigure{\includegraphics[height=160mm]{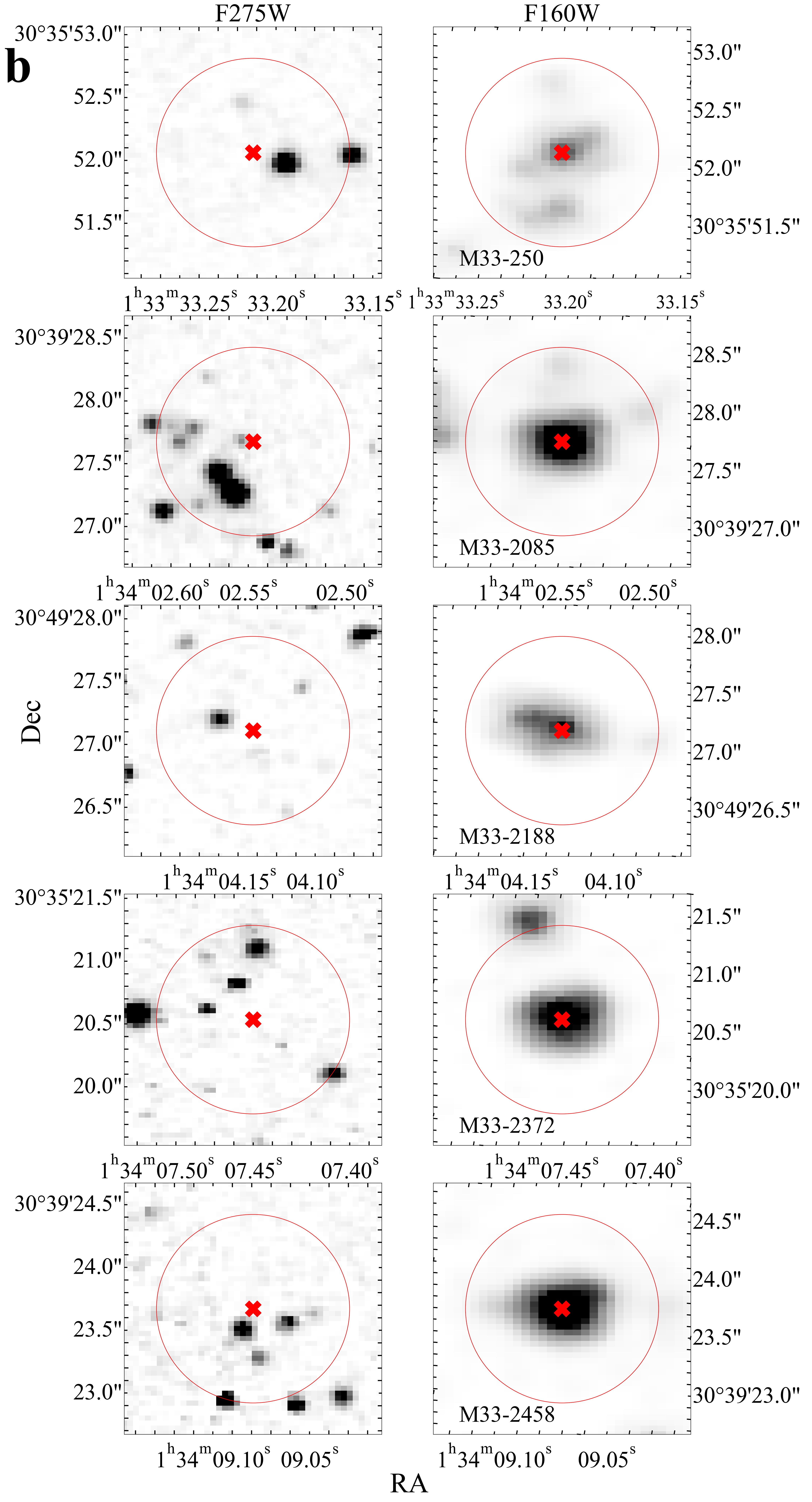}}
	\subfigure{\includegraphics[width=85mm]{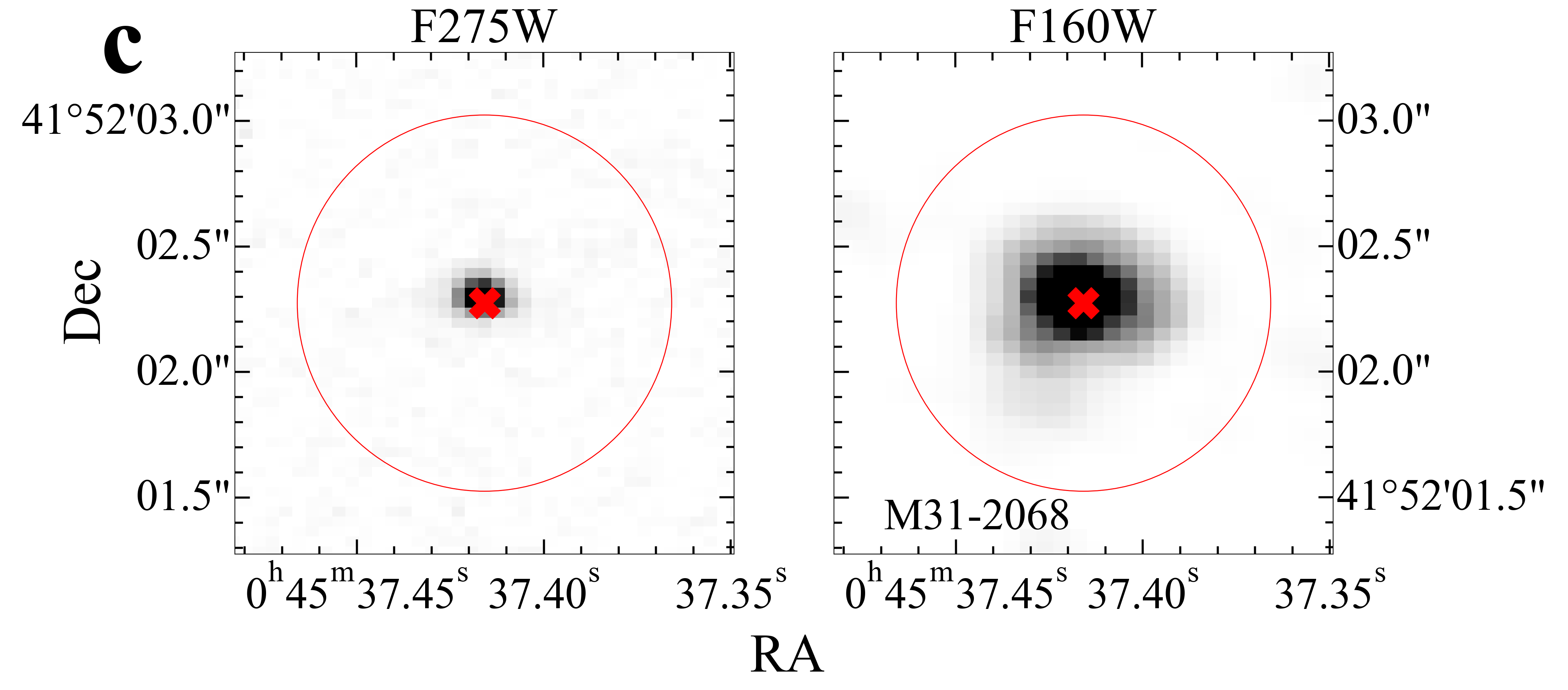}}
	\caption{The mosaic image of 11 stars.}
	\label{fig: DRZ_Mosaics}
\end{figure*}

\citet{2021ApJ...908...87N} also provided a catalog of 182 RSGs with spectral observations and their classifications in M31 and M33. Their binary identification method was based on the detection of blue star features in the spectra of RSGs, as described in \citet{2018AJ....156..225N}. We cross-matched our sample with theirs using a radius of 0.5$^{\prime\prime}$, resulting in 18 and 15 common sources in M31 and M33, respectively. Among these, the binary classification is discrepant for 11 objects. Ten were identified as binaries by \citet{2021ApJ...908...87N} but as single stars in our analysis. This discrepancy is likely due to contamination within the spectroscopic aperture of Hectospec, the Fiber-Fed spectrograph used in their study, which has a diameter of 1.5$^{\prime\prime}$ \citep{2005PASP..117.1411F}. In contrast, the PSF FWHM of the HST is significantly smaller (e.g., $\sim$ 0.08$''$ in the F275W band), allowing better spatial resolution. 
To illustrate this, we present HST F275W and F160W mosaics images of the conflicting sources in panels (a) and (b) of Figure~\ref{fig: DRZ_Mosaics}. Red `X' symbols mark the central positions of the sources, while red ellipses indicate the 1.5$''$ spectroscopic aperture. In each panel, the left column displays the F275W image, and the right column shows the F160W image. In the F275W images, at least one contaminating star is present within the 1.5$''$ aperture, but no source is detected at the center itself, which explains why our work identifies them as single stars, suggesting that the contaminating star affected the spectroscopic classification. Conversely, we identified one star as a binary that was classified as a single star by \citet{2021ApJ...908...87N}. Panel (c) of Figure~\ref{fig: DRZ_Mosaics} shows that this star is clearly detected at the central position in the F275W image, confirming our classification. However, its brightness in F275W is 21.142 mag, which may be too faint to be detected spectroscopically.

\subsection{UV Excess in the Color-color Diagrams} \label{subsec:Color-color_diagrams}
\begin{figure*}[!htb]
    \centering
    \includegraphics[width=170mm]{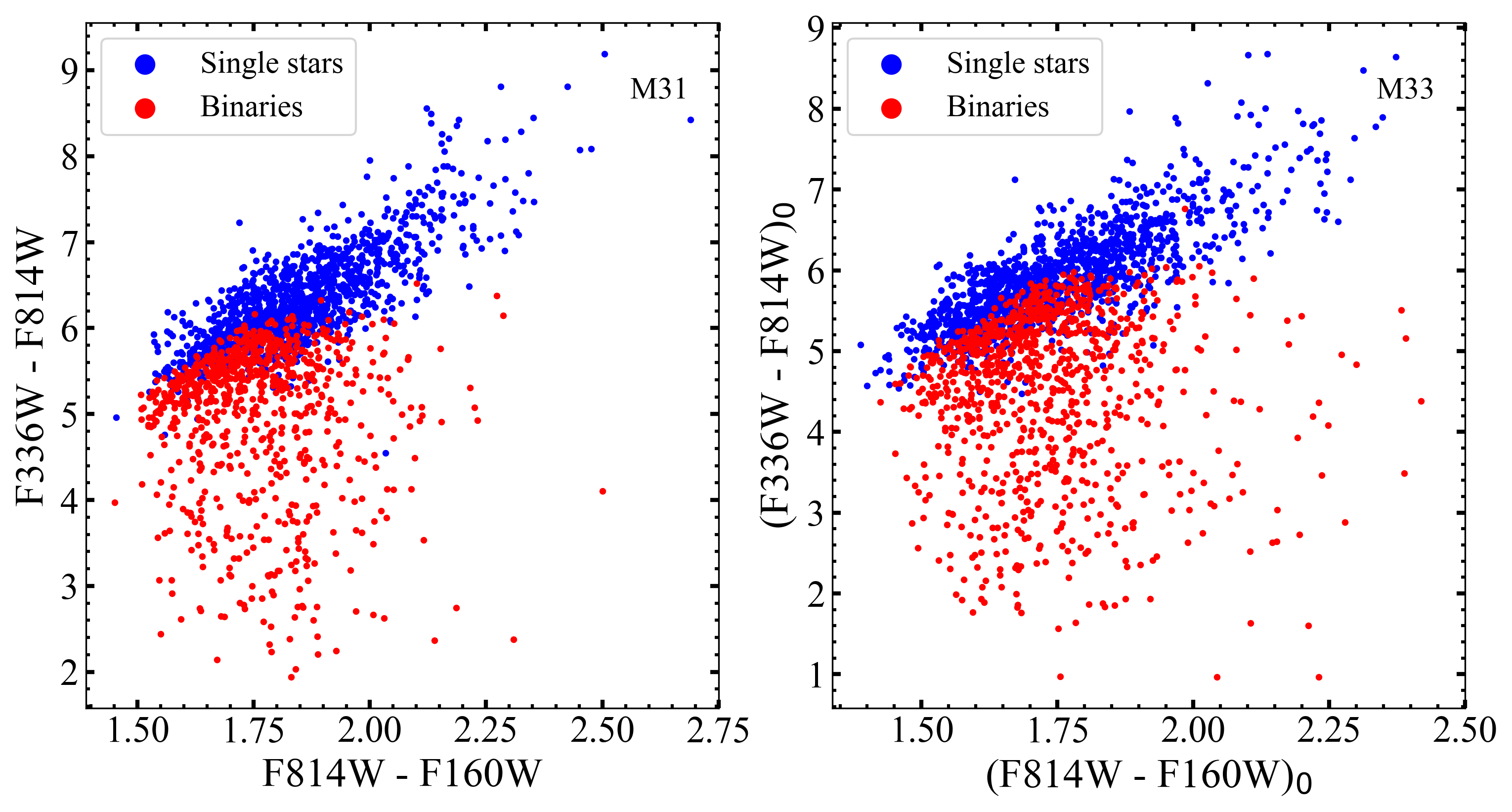}
	\caption{The F336W $-$ F814W versus F814W $-$ F160W diagram for RSGs sample of this work. The red and blue dots represent the RSG binaries and single RSGs identified by this work, respectively.}
    \label{fig: CCD_F336W-F814Wvs.F814W-F160W}
\end{figure*}

Theoretically, RSG binaries are expected to exhibit larger F336W $-$ F814W colors than single stars, due to the additional contribution from their companion stars in the UV bands. Consequently, RSG binaries are anticipated to occupy the lower-right region of the F336W $-$ F814W versus F814W $-$ F160W diagram. Figure \ref{fig: CCD_F336W-F814Wvs.F814W-F160W} shows the distributions of the RSGs in M31 (left) and M33 (right) in the F336W $-$ F814W versus F814W $-$ F160W diagram, where the color indices are corrected for the Galactic foreground extinction, and for M33, additionally for its internal extinction. The red dots represent the binaries identified in this work, and the blue dots represent single stars. It is clear that most stars are located in the areas as expected for single stars and binaries. It should be noted that some binaries (red dots) with F814W $-$ F160W $<$ 2.0 mag are interspersed among the single stars. This can be attributed to the relatively modest UV contribution from some high-temperature companions, which may not produce a significant F336W excess.

\subsection{Binary Fraction of Red Supergiants as a Function of Metallicity?} \label{subsec:galaxies}
Using the derived binary fractions for RSGs with $L/L_{\odot} > 4.0$ of M31 (31.6\%), M33 (34.7\%) from this work, and those for the LMC (26.6\%) and SMC (26.4\%) from \citetalias{2025MNRAS.539.1220D}, we investigated the potential correlation between the RSG binary fraction and host galaxy metallicity. No significant dependence of the observed RSG binary fractions on metallicity was found across these four galaxies. Moreover, the BPASS models predict that the M31, M33, LMC, and SMC binary fractions are 30.9\%, 38.6\%, 32.9\%, and 38.4\%, respectively. The methodology for calculating the BPASS binary fractions for the LMC and SMC follows the approach outlined in Section \ref{subsec:Comparison_binary_fraction_model}, with modifications to the constraints on $T_{\mathrm{eff}}$ and Z. The revised constraints are specified as follows:
\begin{align}
    \text{For LMC:} & \quad 3400 \mathrm{K} \leq T_{\mathrm{eff}} \leq 4250 \mathrm{K}, \ Z = 0.006,\\
	\text{For SMC:} & \quad 3600 \mathrm{K} \leq T_{\mathrm{eff}} \leq 4600 \mathrm{K}, \ Z = 0.002.
\end{align}
The BPASS model predictions also show no apparent correlation between binary fraction and metallicity. However, we found that the agreement between the observed fractions and model predictions is strongest for M31, followed by M33, and weakest for the LMC and SMC. This discrepancy may arise from two factors. First, the model metallicity for M31 is likely better constrained, whereas the irregular morphologies of the LMC and SMC could introduce additional environmental effects on binary fractions. Second, the BPASS model used in this study assumes a fixed IMF, whereas the IMF itself may vary with metallicity and cosmic time \citep{2023Natur.613..460L}. Such variations could contribute to the discrepancies between our observational results and the BPASS predictions. Indeed, \citet{2021ApJ...908...87N} revealed that the binary fraction in M33 exhibits a metallicity dependence. However, there is no direct comparison between the limited metallicity gradient observed in M33 (e.g., \citeauthor{2007A&A...470..865M} \citeyear{2007A&A...470..865M}; \citeauthor{2025arXiv250418779L} \citeyear{2025arXiv250418779L}) and the galaxy-to-galaxy metallicity variation discussed in our earlier analysis. Furthermore, our current sample of RSGs in M33 is located within the inner regions, preventing a comprehensive investigation of this correlation.

\section{Summary} \label{subsec:summary}
In this work, we presented more reliable and complete samples of RSGs and calculated their binary fractions in M31 and M33 using HST archive data. We identified 2,612 and 3,294 RSGs in M31 and M33, respectively, using the F110W $-$ F160W versus F160W color-magnitude diagrams after removing foreground stars based on the F110W $-$ F160W versus F814W $-$ F110W, F110W $-$ F160W versus F475W $-$ F110W, and F814W $-$ F160W versus F475W $-$ F814W color-color diagrams. Accounting for the coverage of PHAT and PHATTER surveys, the total numbers of RSGs in M31 and M33 galaxies are estimated to be 6,563 and 7,572, respectively. Identification of RSG binary candidates was based on their observed flux significantly exceeding that predicted by a single-RSG model in the F275W and F336W bands. The RSG binary fractions are determined to be 33.4\% $\pm$ 0.9\% and 30.9\% $\pm$ 0.8\% for the full samples in M31 and M33, respectively, and 31.6\% $\pm$ 1.9\% and 34.7\% $\pm$ 1.8\% for the high-luminosity subsamples with log $L/L_{\odot} > 4.0$ in M31 and M33, respectively. The derived binary fractions are in good agreement with predictions from the BPASS model. Based on SED fitting of the RSGs, we derived their physical parameters, such as $T_{\mathrm{eff}}$, $R$, and $L$.

\section{Acknowledgments}
We are grateful to the anonymous referee for their very helpful comments and suggestions to improve this paper. We are grateful to Drs. Xiaodian Chen, Jun Li, Yi Ren and Ming Yang for their helpful discussion and suggestion. This work is supported by the NSFC project 12133002 and 12373028, National Key R\&D Program of China No. 2019YFA0405500. S.W. acknowledges the support from the Youth Innovation Promotion Association of the CAS (grant No. 2023065). This work has made use of the data from HST.

\software{Astropy \citep{2022ApJ...935..167A}, Matplotlib \citep{Hunter:2007}, $pandas$ \citep{mckinney-proc-scipy-2010}, $scikit$-$learn$ \citep{scikit-learn}, TOPCAT \citep{2005ASPC..347...29T}.}
\bibliography{binary_m31_m33}{}
\bibliographystyle{aasjournal}


\begin{table*}[ht]
\begin{center}
	\caption{\label{Number of Foreground Stars} Number of Foreground Stars Removed based on the HST and Gaia Observations}
	\vspace{0.1in}
	\hspace{-1.0in}
	\begin{tabular}{cccccccc}
		\hline \hline
		Regions	& Initial Sample& CCD1$^{a}$  & CCD2$^{b}$ & CCD3$^{c}$ & Gaia & Foreground Stars$^{d}$ & Member Stars \\
		\hline
		M31 (PHAT) & 194,400   &  14,544  & 14,092 &  19,658 & 1,905 & 21,020  & 173,380 \\
		M33 (PHATTER) & 110,563  &  9,816  & 11,734 &  9,509 & 86 & 12,574 & 97,989 \\
		\hline
		\multicolumn{8}{l}{\footnotesize{$^a$ F110W $-$ F160W versus F814W $-$ F110W diagram.}}\\
		\multicolumn{8}{l}{\footnotesize{$^b$ F110W $-$ F160W versus F475W $-$ F110W diagram.}}\\
		\multicolumn{8}{l}{\footnotesize{$^c$ F814W $-$ F160W versus F475W $-$ F814W diagram.}}\\
		\multicolumn{8}{l}{\footnotesize{$^d$ all foreground stars.}}
	\end{tabular}   
\end{center}
\end{table*}

\begin{deluxetable}{ccccccc}[ht]\label{tab: M31_results}
	\tablecaption{The derived physical parameters and classification of RSGs in M31}
	\tablewidth{0pt}
	\tablehead{\colhead{No.} & \colhead{R.A. (J2000)} & \colhead{Decl. (J2000)} & \colhead{$T_{\mathrm{eff}}$} & \colhead{$R$} & \colhead{log $L$} & \colhead{Type$^{a}$}\\
		\colhead{} & \colhead{deg} & \colhead{deg} & \colhead{K} & \colhead{$R_{\odot}$} & \colhead{$L_{\odot}$} & \colhead{}}
	\startdata
	M31-1& 10.5911765& 41.2470391& 3800&  148& 3.62& S     \\
	M31-2& 10.5928188& 41.2675842& 3700 & 144& 3.54& S    \\  
	... & ... & ...  & ... & ...&...&... \\
	M31-2611& 11.9133183& 42.1131302& 3650&  304& 4.17& B   \\ 
	M31-2612& 11.9312454& 42.1765637& 3700&  133& 3.48& S     
	\enddata
	\tablecomments{$^{a}$ S = single star, B = binary system. `-' represents the reduced $\chi^{2}$ of the SED fitting greater than 95\% confidence level. \\
		(This table is available in its entirety in machine-readable form.)}
\end{deluxetable}

\begin{deluxetable}{ccccccc}[ht]\label{tab: M33_results}
	\tablecaption{The derived physical parameters and classification of RSGs in M33}
	\tablewidth{0pt}
	\tablehead{\colhead{No.} & \colhead{R.A. (J2000)} & \colhead{Decl. (J2000)} & \colhead{$T_{\mathrm{eff}}$} & \colhead{$R$} & \colhead{log $L$} & \colhead{Type$^{a}$}\\
		\colhead{} & \colhead{deg} & \colhead{deg} & \colhead{K} & \colhead{$R_{\odot}$} & \colhead{$L_{\odot}$} & \colhead{}}
		\startdata
		M33-1& 23.3438331& 30.5271759& 3650&  206& 3.83& S     \\
		M33-2& 23.3461308& 30.5288034& 3950&  150& 3.69& S     \\
		... & ... & ...  & ... & ...&...&... \\
		M33-3293& 23.6549810& 30.7859511& 4050&  124& 3.57& S     \\
		M33-3294& 23.6551653& 30.7854968& 3850&  116& 3.43& S   
		\enddata
		\tablecomments{$^{a}$ S = single star, B = binary system. `-' represents the reduced $\chi^{2}$ of the SED fitting greater than 95\% confidence level.  \\
			(This table is available in its entirety in machine-readable form.)}
\end{deluxetable}

\end{CJK}
\end{document}